\documentclass[a4paper,12pt]{article}

\usepackage{geometry}
\usepackage[english]{babel}

\usepackage{amsmath}
\usepackage{amsfonts}
\usepackage{amstext}
\usepackage{amssymb}
\usepackage{amsthm}
\usepackage{amscd}
\usepackage{mathrsfs}

\usepackage{xcolor}
\definecolor{myurlcolor}{rgb}{0,0.5,0}
\definecolor{mycitecolor}{rgb}{0,0,1}
\definecolor{myrefcolor}{rgb}{0.5,0,0}
\usepackage{graphicx}
\usepackage{tikz}
\usepackage{tikz-cd}

\usepackage[pagebackref,draft=false]{hyperref}
\hypersetup{colorlinks,
linkcolor=myrefcolor,
citecolor=mycitecolor,
urlcolor=myurlcolor}

\usepackage[capitalize]{cleveref}
\usepackage{caption}

\def\beqa{\begin{eqnarray}}
\def\eeqa{\end{eqnarray}}
\def\bean{\begin{eqnarray*}}
\def\eean{\end{eqnarray*}}

\newcommand{\be}{\begin{equation}}
\newcommand{\ee}{\end{equation}}
\newcommand{\bea}{\begin{eqnarray}}
\newcommand{\eea}{\end{eqnarray}}

\newcommand{\lag}{\mathscr{L}}

\newcommand*{\Hrulefillsubpar}[1][0.4pt]{%
  \leavevmode\leaders\hrule height #1\hfill\kern0pt}
\newcommand*{\Hrulefillpar}[1][0.8pt]{%
  \leavevmode\leaders\hrule height #1\hfill\kern0pt}
\newcommand*{\Hrulefillsec}[1][1.6pt]{%
  \leavevmode\leaders\hrule height #1\hfill\kern0pt}
\newcommand*{\Hrulefillsubsec}[1][1.2pt]{%
  \leavevmode\leaders\hrule height #1\hfill\kern0pt}

  \numberwithin{equation}{section}

\usepackage[pagebackref,draft=false]{hyperref}
\hypersetup{colorlinks,
linkcolor=myrefcolor,
citecolor=mycitecolor,
urlcolor=myurlcolor}

\usepackage[capitalize]{cleveref}
\usepackage{caption}
\usepackage{etaremune}

\renewenvironment{thebibliography}[1]
         {\section*{References}\frenchspacing\small
          \begin{list}{[\arabic{enumi}]}
         {\usecounter{enumi}\parsep=2pt\topsep 0pt
         \settowidth{\labelwidth}{[#1]}
         \leftmargin=\labelwidth\advance\leftmargin\labelsep
         \rightmargin=0pt\itemsep=1pt\sloppy}}{\end{list}}

\input xy
\xyoption{all}

\newcommand{\eqn}[1]{(\ref{#1})}

\newtheorem{remark}{Remark}

\newtheorem{example}{Example}

\date{}

\author{F.M.Ciaglia$^{1}$ \href{https://orcid.org/0000-0002-8987-1181}{\includegraphics[scale=0.7]{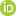}}, F.Di Cosmo$^{2,3}$ \href{https://orcid.org/0000-0003-0256-5913}{\includegraphics[scale=0.7]{ORCID.png}}, A.Figueroa$^{4}$ \href{https://orcid.org/0000-0002-8131-8883}{\includegraphics[scale=0.7]{ORCID.png}}, V.I.Man'ko$^{5,6}$\\ G.Marmo$^{4,7}$ \href{https://orcid.org/0000-0003-2662-2193}{\includegraphics[scale=0.7]{ORCID.png}}, L.Schiavone$^{8}$ \href{https://orcid.org/0000-0002-1817-5752}{\includegraphics[scale=0.7]{ORCID.png}}, F.Ventriglia$^{4,7}$, P.Vitale$^{4,7}$  \href{https://orcid.org/0000-0002-5146-410X}{\includegraphics[scale=0.7]{ORCID.png}} \\
\footnotesize{$^1$ Max-Planck-Institut f\"{u}r Mathematik in den Naturwissenschaften}\\ \footnotesize{Inselstra{\ss}e 22, 04103 Leipzig, Germany.} \\
\footnotesize{$^2$ Dep.to de Matematica, Univ. Carlos III de Madrid}\\\footnotesize{ Av.da de la Universidad, 30, 28911, Leganes, Madrid, Spain.}\\
\footnotesize{$^3$ICMAT, Instituto de Ciencias Matematicas (CSIC-UAM-UC3M-UCM)}\\ \footnotesize{Nicolás Cabrera,1315, Campus de Cantoblanco, UAM, 28049, Madrid, Spain.}\\
\footnotesize{$^4$ INFN-Sezione di Napoli, Complesso Universitario di Monte S. Angelo} \\ \footnotesize{Edificio 6, via Cintia, 80126 Napoli, Italy.} \\
\footnotesize{$^5$ Lebedev Physical Institute, Leninskii Prospect 53, Moscow 119991, Russia.}\\
\footnotesize{$^6$Moscow Institute of Physics and Technology (State University)} \\ \footnotesize{Dolgoprudny, Moscow Region, Russia.} \\
\footnotesize{$^7$ Dipartimento di Fisica ``E. Pancini'', Universit\`a di Napoli Federico II} \\ \footnotesize{Complesso Universitario di Monte S. Angelo Edificio 6, via Cintia, 80126 Napoli, Italy.}\\
\footnotesize{$^8$ Department of Mathematics, Faculty of Science, University of Ostrava, 30}\\ \footnotesize{Dubna 22, 701 03  Ostrava, Czech Republic.}}

\begin{document}



\title{Nonlinear Dynamics from Linear Quantum Evolutions}



\maketitle

\begin{abstract}
Linear dynamics restricted to invariant submanifolds generally gives rise to nonlinear dynamics. Submanifolds in the quantum framework may emerge for several reasons: one could be interested in specific properties possessed by a given family of states, either as a consequence of experimental constraints or inside an approximation scheme. In this work we investigate such  issues in connection with a one parameter group $\phi_t$ of transformations on a Hilbert space, $\mathcal{H}$, defining the unitary evolutions of a chosen quantum system. Two procedures will be presented: the first one consists in the restriction of the vector field associated with the Schr\"{o}dinger equation to a submanifold invariant under the flow $\phi_t$. The second one makes use of the Lagrangian formalism and can be extended also to non-invariant submanifolds, even if in such a case the resulting dynamics is only an approximation of the flow $\phi_t$. Such a result, therefore, should be conceived as a generalization of the variational method already employed for stationary problems.
\end{abstract}



\section{Introduction}
\label{sec-Introduction}

The dynamical evolution of a quantum non-relativistic system is described by means of  the {Schr\"odinger equation}
\be \label{Eq: schrodinger equation}
i\frac{d}{dt} \psi \,=\, {\mathbf{H}} \psi\,,
\ee
where $\psi$ is an element of the Hilbert space associated with the system, say $\mathcal{H}$, and ${\mathbf{H}}$ is an Hermitian operator on $\mathcal{H}$. Equation \eqn{Eq: schrodinger equation} is a linear, first order ordinary differential equation, and thus may be described by means of a linear vector field on  $\mathcal{H}$.
On the other hand, the dynamical evolution of classical systems is described  by means of vector fields which need not be linear, and this seems to be somehow strange in view of Dirac's idea that Classical Mechanics should emerge as a suitable limit of Quantum Mechanics. 

In this respect, quite a few years ago, some of us have shown how nonlinear interacting systems could be obtained starting with free linear systems \cite{Manko_Marmo_Generalized_reduction_procedure_and_nonlinear_non-stationary_dynamical_systems}. 
The main idea advanced there was that, if we restrict the family of allowed initial conditions for some dynamics to a suitable invariant nonlinear submanifold, the ``restricted'' dynamics turns out to be nonlinear. This is surprising because linear structures in Physics generally arise as approximations to the more accurate nonlinear ones. What we face here is at variance with our expectations: it is the deeper and more complete theory that is linear and the nonlinear classical (approximate) theory is to arise as a suitable limit. Therefore, one of the aim of this paper is to clarify what are the essential ingredients which permit to obtain non-linear dynamics in the quantum setting, and what is the relation between the original linear dynamiccs and the induced nonlinear one.

In this paper we would like to consider this situation in the framework of unitary evolution maps on a given Hilbert space associated with a quantum system.
In this context, we will consider nonlinear submanifolds of $\mathcal{H}$ because, for instance, they exhibit a behaviour as close as possible to classical states (coherent states and their generalizations), or as a consequence of some prescribed limitations (experimental setups, approximation procedures). While submanifolds deriving from an experimental constraint have to be chosen case-by-case via ad-hoc procedures, there are more general constructions which lead to the introduction of systems of coherent states. One of the first possibilities which has been investigated consists in the definition of suitable {\it displacement operators} (see \cite{Ciaglia_DiCosmo_Ibort_Marmo_dynamical_aspects_in_the_quantizer_dequantizer_formalism} for more details about the use of quantizer-dequantizer formalism for the definition of generalized coherent states). 
What are nowadays called { \it Weyl systems} were introduced by H. Weyl to deal with unbounded operators in the canonical commutation relations.
The infinitesimal version, given in terms of Hermitean operators satisfying  canonical commutation relations $[\, \mathbf{q}\,,\mathbf{p} \,]=\imath\hbar\, \mathbb{I}$, was replaced by an ``exponentiated version'' given in terms of unitary operators on a Hilbert space, $\mathcal{H}$:
\be \label{Eq: weyl systems}
\mathbf{W}(a_{1})\,\mathbf{W}(a_{2})\mathbf{W}^{\dagger}(a_{1})\,\mathbf{W}^{\dagger}(a_{2}) \,=\, \mathrm{e}^{\imath \, \omega(a_{1}\,,a_{2})} \, \mathbb{I}\,,
\ee
where $a_{1},a_{2}$ are elements of an Abelian vector group $\mathbb{V}$ endowed with a translation invariant symplectic structure $\omega$. The map $\mathbf{W}$ from $\mathbb{V}$ to $\mathcal{U}(\mathcal{H})$ is called a  {\it Weyl system}.
As it is clear from the above relation, a Weyl system defines a projective unitary representation of the Abelian group $\mathbb{V}$ over $\mathcal{H}$. The action of these operators on a fiducial state $\psi_0$ produces an immersion, say $\mathfrak{i}$, of the classical-like phase space $\mathbb{V}$ into the quantum Hilbert space. Indeed, we can write the following map:
\be \label{Eq: immersion phase space into hilbert space}
\mathfrak{i} \;\;|\;\; \mathbb{V} \to \mathcal{H} \;\;:\;\; v \mapsto \mathfrak{i}(v) \,=\, \mathbf{W}(v) \,\psi_0 \,=:\, \psi_v\,,
\ee
which is an injection of $\mathbb{V}$ into $\mathcal{H}$. 
Moreover, being $\mathbf{W}(v)$ a unitary operator, $\psi_v$ lies in a homogeneous space with respect to the action of the unitary group $\mathcal{U}(\mathcal{H})$. 
Therefore, starting with a normalized vector lying on the unit sphere $\mathcal{S}(\mathcal{H})$, equation \eqn{Eq: immersion phase space into hilbert space} defines an immersion of $\mathbb{V}$ into $\mathcal{S}(\mathcal{H})$ which is, again, injective. 
However, it is worth noticing that even though $\mathbb{V}$ is a vector space, the immersion does not respect linearity, i.e., $\psi_{v_1} + \psi_{v_2}\neq \psi_{v_{1} + v_{2}}$ in general, and $\mathfrak{i}(\mathbb{V})$ is not a linear subspace of $\mathcal{H}$. 
In summary, generalized coherent states allow us to define injective maps from a given set to the Hilbert space $\mathcal{H}$ associated with a quantum system, every point of the set labelling a given vector in $\mathcal{H}$. 
This is the main property we will exploit in the first part of the paper in order to study how to induce one parameter groups of transformations on subsets of states starting from unitary maps on $\mathcal{H}$.  
Using coherent states it is possible to interpret these induced maps as classical-like dynamics in the framework of classical-to quantum transition (see \cite{Ciaglia_DiCosmo_Ibort_Marmo_dynamical_aspects_in_the_quantizer_dequantizer_formalism, almalki_kisil-geometric_dynamics_of_a_harmonic_oscillator,almalki_kisil-solving_the_schrodinger_equation_by_reduction}). However, it is worth stressing that the following discussion can be extended to submanifolds which do not possess classical-like properties, representing, therefore, generic constraints. Moreover, we do not require our subset of states to possess any completeness property, i.e., to provide a "resolution of the identity".

In the forthcoming sections, we will consider two different procedures to achieve such a result. 
Let us shortly present their main features in this introduction, their complete descriptions constituting the main body of the paper. Let us start with the subset $\mathfrak{i}(\mathbb{V}) \subset \mathcal{H}$ of parameterized states. 
Let $\phi_t : \mathcal{H} \rightarrow \mathcal{H}$ be the one-parameter group of unitary transformations generated by ${\mathbf{H}}$ according to the Schr\"{o}dinger equation \eqref{Eq: schrodinger equation}. If the submanifold $\mathfrak{i}(\mathbb{V})$ is invariant with respect to this flow, i.e., if it holds
\be \label{Eq: invariant submanifold}
\forall \,\, t \in \mathbb{R} \;\;\text{ and } \;\; \forall \,\,v \in \mathbb{V}\,, \;\;\; \exists \, w \in \mathbb{V} \;\; : \;\; \phi_t \left(\psi_v \right) \,=\, \psi_w ,
\ee
it is possible to induce a dynamics on $\mathbb{V}$ according to the following commutative diagram
\begin{center}
\begin{tikzcd}
\mathbb{V} \arrow[r, hook, "\mathfrak{i}"] \arrow[d,"\tilde{\phi}_t"] & \mathcal{S}(\mathcal{H}) \arrow[d, "\phi_t"] \\
\mathbb{V} \arrow[r, hook, "\mathfrak{i}"] & \mathcal{S}(\mathcal{H}) .
\end{tikzcd}
\end{center}
Essentially, a flow on $\mathbb{V}$, say $\tilde{\phi}_t$, is defined in such a way that
\be \label{Eq: induced dynamics invariant submanifold}
\mathfrak{i} \cdot \phi_t \,=\, \tilde{\phi}_t \cdot \mathfrak{i} .
\ee
It is worth stressing the fact that, even though  quantum dynamics described by $\phi_t$ is a linear one, it does not necessarily mean that $\tilde{\phi}_t$ is linear too.
Indeed, we will see explicit examples in which $\tilde{\phi}_t$ is associated with a nonlinear dynamics. 

It is quite clear that we may replace the vector space $\mathbb{V}$ in the previous construction  with a generic differential manifold, say $\mathcal{M}$,  and the immersion \eqn{Eq: immersion phase space into hilbert space} with an injection of $\mathcal{M}$ onto an invariant submanifold in $\mathcal{H}$ or $\mathcal{S}(\mathcal{H})$. 
These ingredients, again, will allow us to induce a one parameter group of transformations $\tilde{\phi}_t$ on $\mathcal{M}$, which we can interpret as an evolution map and we would expect it to be nonlinear, in general. 
From the physical point of view, $\mathbb{V}$ has a direct interpretation as a ``classical phase-space'', while, in general, the manifold $\mathcal{M}$ may not have  a preferred physical interpretation. 
However, in all the examples we will propose, we will always choose $\mathcal{M}$ to give rise to an immersed submanifold, $\mathfrak{i}(\mathcal{M})$, of states exhibiting a semi-classical behaviour, for instance the so called  {\it squeezed and correlated states} which we will address in section \eqn{Sec: squeezed correlated states}.  

In the procedure outlined above, the crucial assumption is expressed by equation \eqn{Eq: invariant submanifold}, namely, we have to assume the immersed submanifold $\mathfrak{i}(\mathbb{V})$ to be invariant under the flow $\phi_t$. 
Geometrically speaking, this is equivalent to have the linear vector field generating the Schr\"{o}dinger equation to be tangent to the immersed submanifold.
In the language of differential geometry, we may describe this procedure to be contravariant-like in the sense that it operates by means of restriction of contravariant objects, i.e., vector fields.
Clearly, there is no a-priori reason for the vector field generating Schr\"{o}dinger equation to be tangent to the immersed submanifold, and a case-by-case analysis is required to verify this statement.
In this context,  we are not free to choose the immersed submanifold if we fix the quantum dynamics, and this may be thought of as a limitation to this contravariant-like procedure.

On the other hand, having at our disposal the immersion map 
\be
\mathfrak{i} \;\; |\;\; \mathcal{M} \hookrightarrow \mathcal{H}
\ee
of $\mathcal{M}$ into $\mathcal{H}$, we are always in the position of considering the pullback to $\mathcal{M}$ of covariant tensors on $\mathcal{H}$, and this instance brings in the possibility of defining a covariant-like procedure for inducing a dynamics on $\mathcal{M}$ starting with a Lagrangian description for the Schr\"{o}dinger equation \eqn{Eq: schrodinger equation} on $\mathcal{H}$.
Indeed, since the Lagrangian formalism is a description of the dynamics which is based on the use of differential forms (covariant tensors), one could exploit a variational formulation in terms of a Lagrangian function to induce a one parameter group of transformations also in the case of  non-invariant submanifolds. 
Starting from  a suitable Lagrangian function on the tangent bundle $\textbf{T}\mathcal{H}$ of $\mathcal{H}$, equation \eqn{Eq: schrodinger equation} may be written in terms of a vector field $\Gamma_{l}$, which is the lift to $\textbf{T}\mathcal{H}$ of the vector field $\Gamma$ on $\mathcal{H}$ describing the Schr\"{o}dinger equation, satisfying 
\be 
i_{\Gamma_{l}} \, \omega_\lag \,=\, -\mathrm{d}E_\lag\,,
\ee 
and even though $\Gamma$ is not tangent to $\mathcal{M}$, it is still possible to pullback to the tangent bundle of $\mathcal{M}$ the two-form $\omega_{\lag}$ and the Lagrangian energy function $E_{\lag}$.
Then, if the pull-back of $\omega_{\lag}$ turns out to be non-degenerate\footnote{When the pulled-back Lagrangian  gives rise to a degenerate 2-form,  an evolution on the tangent bundle of $\mathcal{M}$ can be obtained only after performing a constraints analysis according to the Dirac-Bergmann procedure.}, we may define a vector field $\tilde{\Gamma_{l}}$ on $\textbf{T}\mathcal{M}$   by means of   
\be
i_{\tilde{\Gamma_{l}}} \tilde{\omega}_\lag \,=\, - \tilde{\mathrm{d}}\tilde{E}_\lag\,, 
\ee
where 
\be
\begin{split}
\tilde{\omega}_\lag & \,=\, (T\mathfrak{i})^* \omega_\lag\,, \\
\tilde{E}_\lag & \,=\, (T\mathfrak{i})^* E_\lag\,, 
\end{split}
\ee 
and $\tilde{\mathrm{d}}$ is the differential on $\mathcal{M}$, and this vector field generates a dynamical evolution on $\textbf{T}\mathcal{M}$ which is, in general, nonlinear. 
Moreover, we will see that the integral curves of $\tilde{\Gamma_{l}}$ may be projected down on $\mathcal{M}$, thus generating a dynamical evolution on $\mathcal{M}$.
Therefore, differently from what we found in the case of the contravariant-like procedure described above (see equation \eqn{Eq: induced dynamics invariant submanifold}), this covariant-like procedure just explained is always possible, even  if the immersed submanifold is not invariant under the unitary dynamics on the ambient Hilbert space. 

If $\mathcal{M}$ is invariant under the unitary dynamics, we may perform both the procedures, and we will see that the resulting dynamical evolutions coincide on $\mathcal{M}$.
However, \textbf{the induced dynamics on a non-invariant submanifold can no longer be seen as a restriction of the quantum unitary evolution on the Hilbert space, and the induced Lagrangian description may be conceived as an approximation of the quantum one  generalizing the approximantion scheme known as variational method.} 
In this framework, indeed, one selects a submanifold of states and the minimum of the energy functional restricted to this submanifold, if any, is the best approximation of the ground state of the Hamiltonian operator on this submanifold. 
The situation we are dealing with, instead, will correspond to the approximation of a dynamical evolution.   

Let us now illustrate the plan of the work. It is divided into two parts: in the first two sections we focus on induced dynamics on invariant submanifolds. In particular, in section \ref{Sec: squeezed correlated states}, we consider the restriction of the Hamiltonian operators defining the free motion of a system with a single degree of freedom and a one-dimensional harmonic oscillator to the invariant submanifold $\mathcal{M}$ of Gaussian states.
Both restrictions give rise to nonlinear dynamics on $\mathcal{M}$. In section \ref{sec: coherent states and nonlinear embeddings}, we  consider nonlinear immersions in terms of coherent states, the corresponding submanifolds still being invariant for the chosen one parameter groups of transformations on the image Hilbert space $\mathcal{H}$. 
In this case, we present examples where the parameter manifolds are linear spaces, the immersion maps are nonlinear (in the sense that the immersed submanifolds are not linear subspace of the ambient Hilbert space), and the induced dynamical evolutions on the parameter submanifolds are actually linear with respect to the linear structure of the parameter manifolds. 
This clearly highlights the fact that the nonlinearity of the immersion and the nonlinearity of the induced flow are not related to each other.

The second part consists of section \ref{Sec: lagrangian formalism} and it is devoted to the description of the Lagrangian procedure previously sketched. 
We firstly introduce a variational principle for the Schr\"{o}dinger equation in terms of a Lagrangian function on the tangent bundle $\mathbf{T}\mathcal{H}$ of the Hilbert space. 
This Lagrangian function is degenerate and has some invariance properties which are consequence of the probabilistic intepretation of quantum mechanics, and one has to define a suitable connection in order to correctly recover the Schr\"{o}dinger equation. 
For the sake of clarity, in the rest of  section \ref{Sec: lagrangian formalism}, we illustrate a series of examples. In particular, we reformulate the examples of section \ref{Sec: squeezed correlated states} according to the variational perspective, showing that this Lagrangian formalism provides the same results as the procedure used in section \ref{Sec: squeezed correlated states} when the submanifold is invariant and, in addition, shows that the derived equations are actualy Hamiltonian ones. The third example focuses on a non-invariant case, namely the submanifold of Gaussian states with respect to the evolution generated by the Hamiltonian operator of a one dimensional anharmonic oscillator. The final example, instead, consists of the immersion of a supermanifold in the Hilbert space of a finite dimensional Hilbert space, and the resulting dynamical system is described by means of a Lagrangian superfunction. Therefore, this example shows that covariant-like procedure proposed in section \ref{Sec: lagrangian formalism} is general enough and may be used also for Fermionic systems. 

A final section contains some remarks on possible generalizations and future perspectives.

\section{Reduction of squeezed and correlated states evolution} \label{Sec: squeezed correlated states}

In this section we consider  unitary evolutions  in the Hilbert space $\mathfrak{L}^2(\mathbb{R}, \, \mathrm{d}x)$ generated by an Hamiltonian operator ${\mathbf{H}}$. In particular, we consider the Hamiltonian operators describing the free motion and the harmonic oscillator. As immersed submanifold of states, we consider the submanifolds of {\it Gaussian states} because they are a particular subset of states where {\it classical-like} properties may be easily identified. Moreover, as the sum of two Gaussian functions is not a Gaussian function, the submanifold is not a vector subspace of the ambient Hilbert space. Specifically, we will consider  {squeezed and correlated states} of the harmonic oscillator \cite{Dodonov_Kurmyshev_Manko_Generalized_uncertainty_relation_and_correleted_coherent_states}; they satisfy the  {Schr\"odinger-Robertson \cite{Robertson_the_uncertainty_principle,Schrodinger_Zum_Heisenbergschen_Unscharfeprinzip} uncertainty relation} minimum. 
These Gaussian states are obtained by means of the following immersion:
\be \label{Eq: immersion squeezed correlated states}
\mathfrak{i} \;\; |\;\; \mathbb{C}_{r+}  \times \mathbb{C}^2 \hookrightarrow \mathfrak{L}^2(\mathbb{R},\, \mathrm{d}x) \;\;:\;\; (a,\,b,\,c) \mapsto \psi(a,\,b,\,c;\,x) \,=\, e^{-a x^2 + b x + c}\,,
\ee
where $\mathbb{C}_{r+}$ denotes the open subset of the complex plane made up of complex number with strictly positive real part.

We will see that the submanifold of $\mathfrak{L}^2(\mathbb{R},\, \mathrm{d}x)$ selected by the immersion \eqn{Eq: immersion squeezed correlated states} is an invariant one under the evolution generated by the following Hamiltonian operators\footnote{We will put always $\hbar \,=\, m \,=\, 1$.}
\be
{\mathbf{H}}_{free} \,=\, -\frac{1}{2} \Delta \qquad {\mathbf{H}}_{harm}\,=\, -\frac{1}{2}(\Delta - \omega^2 x^2) \,\,. 
\ee
Indeed, in both cases, evolving an initial state
\be
\psi(x; \,t=0)\,=\, e^{-a x^2 + bx + c}\,,
\ee
of the type given in equation \eqref{Eq: immersion squeezed correlated states} by means of  the {Green function}
\be
\psi(x,\,t) \,=\, \int_{\mathbb{R}} G(x,\,x',\,t) \,\psi(x,\,t) \,\mathrm{d}x' ,
\ee
one gets again a state of the type
\be \label{Eq: squeezed correlated states}
\psi(x,\,t) \,=\, e^{-a(t) x^2 + b(t)x + c(t)} \,\,,
\ee
where $a(0)= a$, $b(0)= b$, $c(0)= c$, and, if the initial data satisfy $\Vert \psi(0)\Vert^2 \,=\,1$, then $a(t),\,b(t),\,c(t)$ will be such that $\Vert \psi(t) \Vert^2 \,=\,1 \;\; \forall \,t \in \mathbb{R}$. Therefore, the dynamical vector field is tangent to the hyper-surfaces of constant norm.
In order to obtain the dynamics $\tilde{\phi}_t$ on $\mathcal{M}\,=\, \mathbb{C}_{r+}\times \mathbb{C}^2$, namely, the evolutionary equations for $a,\,b,\,c$ out of the dynamics $\phi_t$ associated with $\hat{\mathscr{H}}$, one should solve  {Schr\"odinger equation} for a wave function of the type \eqref{Eq: squeezed correlated states}\footnote{From the geometrical point of view, this  amounts to restrict the Schr\"odinger vector field to the submanifold of squeezed and correlated states which is invariant with respect to the flow associated with the equations of motion.}
\be
i\frac{d}{dt} e^{-a(t) x^2 + b(t)x + c(t)} \,=\, {\mathbf{H}} e^{-a(t)x^2 + b(t)x + c(t)} \,\,.
\ee

\subsection{Free particle} \label{Subsec: free particle}

Let us start with the free Hamiltonian 
\be \label{Eq: free hamiltonian}
{\mathbf{H}}_{free}= \frac{\hat{p}^2}{2}= -\frac{1}{2}\Delta \,.
\ee
The corresponding Green function is 
\be
G(x,\,x',\,t)= \frac{1}{\sqrt{2\pi i t }} e^ {i\frac{(x-x')^2}{2t}},
\ee 
which determines the evolution of the initial wave function according to 
\be
\psi(x,t)= \int G(x,\,x',\,t) \, \psi(x';\,t=0) \,\, \mathrm{d} x'
\ee
The evolution of the state vector $|\psi(0)\rangle\rightarrow |\psi(t)\rangle$ is a unitary one. As a consequence,  the normalization of  the wave function is preserved. Therefore, let us choose the wave function $\psi(x,\,0)$ to be a normalized Gaussian wave function of the form
\be
\psi(x;\,t=0) = e^{-a x^2+b x+ c} \,.
\ee
As we said in the introduction to the present section, the result of the evolution of the initial wave function $\psi(0)$ gives the normalized Gaussian function of the form
\be
\psi (x, \,t)
= \int \frac{1}{\sqrt {2\pi i t}}e^{ \, \left[i \frac{(x-x')^2}{2t} -a x'^2 +b x' + c\right]} \, \mathrm{d} x' \,=\, e^{-a(t) x^2 \,+\, b(t)x \,+\, c(t)}.\label{psiX}
\ee
with 
\be
a(t=0)=a,~~~b(t=0)=b,~~~c(t=0)=c\,,
\ee
Evolutionary equations for $a,\,b,\,c$ are deduced  from the Schr\"odinger equation for the ''parent" wave function
\be
i \frac{d}{d t} \, e^{-a(t) x^2 \,+\, b(t) x \,+\, c(t)} = -\frac{1}{2} \Delta \,\, e^{-a(t) x^2 \, + \, b(t) x \,+\, c(t)}\,,
\ee
This equation yields the relation 
\be 
-i \dot a(t) x^2 + i \dot b(t) x +i \dot c(t)= -\frac{1}{2}\left\{\left[-2 a(t) x+b(t)\right]^2 + (-2 a(t))\right\} \,.
\ee
Therefore, we have the dynamical equations
\be
\begin{aligned}
i \dot a(t) &= 2a^2(t), &  a(0) &= a =a_R+ i a_I\,, \\
i \dot b(t) &= 2a(t) b(t), &  b(0) &= b = b_R +i b_I\,, \\
i \dot c(t) &= a(t)-\frac{1}{2} b(t)^2, & c(0) &= c = c_R + i c_I \,.
\end{aligned}
\ee
One sees that
\be
b(t)= \frac{b}{a} a(t)\,,
\ee
and, thus, the ratio $\frac{b(t)}{a(t)}= \frac{b}{a}$ is an  integral of motion for the dynamics. We can rewrite dynamics in real variables 
\be
a= a_R + i a_I \,, ~~~~~ b = b_R + i b_I \,, ~~~~~ c = c_R + i c_I \,.
\ee
Since 
\be 
\dot a_R + i \dot a_I = -2 i (a_R^2 - a_I^2 + 2 i a_R a_I)\,,
\ee
we get 
\be
\dot a_R = 4 a_R a_I, ~~~~ \dot a_I = 2(a_I^2 - a_R^2)\,,
\ee
Also, from
\be 
\dot b_R + i \dot b_I = -2 i (a_R + i a_I)(b_R +  i b_I) = -2 i \left[(a_R b_R - a_I b_I) + i (a_I b_R + a_R b_I)\right]\,,
\ee
we obtain 
\be
\dot b_R = 2(a_I b_R + a_R b_I), ~~~~\dot b_I = 2 (a_I b_I - a_R b_R)\,.
\ee
For the remaining variables, in view of 
\be
c_R + i c_I = \frac{i}{2}(b_R + i b_I)^2 - i (a_R + i a_I)= \frac{i}{2}(b_R^2 - b_I^2 + 2 i b_R b_I) -i (a_R + i b_R)\,,
\ee
we obtain
\be
\dot c_R = a_I - b_R b_I , ~~~~\dot c_I = -a_R + \frac{1}{2}(b_R^2 - b_I^2).
\ee
It is worth stressing the fact that the dynamics induced on $\mathcal{M}$ is a nonlinear one.
The nonlinear dynamics in terms of the introduced notation admits the following  solutions
\beqa
a_R(t)&=& (a_R+4t a_R a_I)\left[(1-2t a_I)^2+ 4 t^2 a_R^2\right]^{-1}\,,\\
a_I(t)&=& \left[a_I-2t (a_R^2- a_I^2)\right]\left[(1-2t a_I)^2+ 4 t^2 a_R^2\right]^{-1}\,,\\
b_R(t)&=&  \frac{1}{a_R^2+ a_I^2} [(b_R a_R+b_I a_I) a_R(t)+ (b_R a_I-b_I a_R) a_I(t)]\,,\\
b_I(t)&=&  \frac{1}{a_R^2+ a_I^2} [ (b_I a_R-b_R a_I) a_R(t) + (b_R a_R+b_I a_I) a_I(t)]\,,\\
c_R(t)&=& c_R(0) + \int_0 ^t \left( a_I(\tau)- b_R(\tau) b_I(\tau) \right) d\tau\,,\\
c_I(t)&=& c_I(0) + \int_0 ^t \left( \frac{1}{2}(b_R^2(\tau)-b_I^2(\tau))-a_R(\tau) \right) d\tau.
\label{eqs:restriction free particle}
\eeqa
The initial data $c_R(0), c_I(0)$ correspond to the choice of a given normalization factor. Thus, evolution of the real variables $ b_R(t), b_I(t), c_R(t), c_I(t)$ is determined by evolution of the variables $a_R(t), a_I(t)$.
We should notice that even though quantum dynamics is the one of the free particle for one degree of freedom, dynamics associated with the subset of Gaussian states is a nonlinear dynamics quite far from the quantum linear one.

\subsection{Harmonic oscillator} \label{Subsec: harmonic oscillator}

Let us now address the case of harmonic oscillator, where the Hamiltonian is
\be \label{Eq: hamiltonian harmonic oscillator}
{\mathbf{H}} \,=\, -\frac{1}{2}(\Delta - \omega^2 x^2) \,.
\ee
Here, the Green function takes the following form
\be \label{Eq: green function harmonic oscillator}
G(x,\, x', \,t)= \sqrt{\frac{\omega}{2\pi i\sin\omega t}} \,\, \mathrm{exp} \,\, \left[\frac{i\mathrm{cotan} \omega t}{2} (x^2+x'^2)-\frac{xx'}{\sin\omega t}\right]. 
\ee
Again, the submanifold of normalized squeezed and correlated states \eqref{Eq: squeezed correlated states} is an invariant one, that is, given an initial state of the type
\be
\psi(x;\,t=0) \,=\, e^{-ax^2 \,+\,b x \,+\, c}\,,
\ee
under the evolution given by \eqref{Eq: green function harmonic oscillator}
\be
\psi(x,\,t)\,=\, \int_{\mathbb{R}} G(x,\,x',\,t) \, \psi(x';\,t=0) \,\, \mathrm{d}x'\,,
\ee
it remains of the same type
\be
\psi(x,\,t)\,=\, e^{-a(t)x^2 \,+\, b(t)x \,+\, c(t)} \,.
\ee
As in the previous example, being the submanifold invariant we can restrict Schr\"odinger vector field on it obtaining the following vector field
\be 
i \frac{d}{d t} \, e^{-a(t) x^2 \,+\, b(t) x \,+\, c(t)} = -\frac{1}{2} (\Delta - \omega^2 x^2) \,\, e^{-a(t) x^2 \, + \, b(t) x \,+\, c(t)} \,.
\ee
It is a matter of direct computation to check that one obtains the following equations determining the evolution of the parameters $a,\, b,\, c$
\be
\begin{aligned}
i \dot a(t)&= 2 a^2(t) - \frac{1}{2} \omega ^2, & a(t=0)&=a\,, \\
i \dot b(t)&= 2 a(t) b(t) , &  b(t=0)&=b\,, \\
i \dot c(t)&=  a(t) - \frac{1}{2}b^2(t), & c(t=0)&=b .
\end{aligned}
\ee
which in real coordinates read
\be
\begin{aligned}
\dot a_R &= 4 a_R a_I, & \dot a_I &= 2(a_I^2 - a_R^2) - \frac{1}{2}\omega^2 \,,\\
\dot b_R &= 2(a_I b_R + a_R b_I), & \dot b_I &= 2 (a_I b_I - a_R b_R) \,,\\
\dot c_R &= a_I - b_R b_I , & \dot c_I &= -a_R + \frac{1}{2}(b_R^2 - b_I^2)\,.
\end{aligned}
\ee
Solutions of these nonlinear equations are
\begin{align}
a(t)&= -i \frac{\omega \mathrm{cotg} \omega t}{2}+ \frac{\omega^2}{\sin^2\omega t} \frac{1}{4\left(a -i \frac{\omega \mathrm{cotg} \omega t}{2}\right)}\,, \\
b(t)&= -i\frac{\omega b}{\sin \omega t}\frac{1}{2\left(a -i \frac{\omega \mathrm{cotg} \omega t}{2}\right)}\,, \\
c(t)&= c+ \ln \left[ \sqrt{\frac{\omega}{2\pi i \sin\omega t}} \sqrt{\frac{\pi}{\left(a -i \frac{\omega \mathrm{cotg} \omega t}{2}\right)}}\right] \,.
\end{align}
We have thus obtained a nonlinear dynamics for the parameters $a,\,b, \,c$. Of course, for $\omega\rightarrow 0$ we get the free motion dynamics.

\section{Coherent states and nonlinear embeddings}\label{sec: coherent states and nonlinear embeddings}

In this section we will briefly recall the construction of some sets $\mathcal{S} \subset \mathcal{H}$ of states, called coherent states, which can be used to define  a map which associates to any point of a manifold $\mathcal{M}$ a vector of $\mathcal{H}$. From our perspective, these maps are examples of nonlinear embeddings, since their images do not form a vector subspace of $\mathcal{H}$, and, in principle, one could expect the emergence of nonlinear dynamics in such a case. Actually, we will see that the nonlinearity of the emergent dynamics is not related to the nonlinearity of the chosen embedding, i.e., we will see that certain particular unitary evolutions on $\mathcal{H}$ will produce  linear dynamical evolutions on the manifold $\mathcal{M}$. 
Besides that, the following examples should make clear that our procedure should not be understood as identifying a classical system associated with a quantum one. Indeed, we shall  consider some quantum systems with a finite dimensional Hilbert space, and it is clear that the classical limit, as usually understood, should be an evolution on a subset of discrete points which will not be the case.    

\subsection{Radcliffe spin-coherent states}
For spin-$\frac{1}{2}$ systems coherent states were introduced by Radcliffe \cite{Radcliffe_some_properties_of_coherent_spin_states} based on the properties of the generators $\mathbf{J}_{+},\mathbf{J}_{-}$ and $\mathbf{J}_{z}$ of the group $SU(2)$.
These generators satisfy the commutation relations:
\be
[\, \mathbf{J}_{z} \, , \, \mathbf{J}_{\pm} ] = \pm \, \mathbf{J}_{\pm} \, ,	\qquad
[\mathbf{J}_{+}\,,\mathbf{J}_{-}]=2\mathbf{J}_{z}\,.
\ee
Upon determining the fiducial state $|-1/2\rangle$ as the state such that $\mathbf{J}_{-}|-1/2\rangle=0\cdot|-1/2\rangle$ and such that $\mathbf{J}_{+}|-1/2\rangle=|1/2\rangle$,  coherent states are defined as:
\be\label{eqn: radcliff pure coherent states}
|z\rangle:=\mathrm{e}^{z\mathbf{J}_{+} - z^{*}\mathbf{J}_{-}}\,|-1/2\rangle = \exp{\left(
\begin{array}{cc}
0 & z \\
-z^* & 0 
\end{array}  
\right)} |-1/2 \rangle = \left(\begin{matrix} \frac{\sin(|z|)}{|z|}\, z \\ \cos(|z|)\end{matrix}\right)\,.
\ee

One can write the complex number $z = \frac{\theta}{2} \mathrm{e}^{\imath\varphi}$ where $\left( \theta , \varphi \right)$ label the unit vector in the three-dimensional Euclidean space:
\be
\mathbf{n} = \left( \, \sin(\theta) \, \cos(\varphi), \, \sin(\theta) \, \sin(\varphi), \,  \cos(\theta) \, \right) \, ,
\ee 
and 
\be
U(g_{\mathbf{n}}) = \mathrm{e}^{z\mathbf{J}_{+} - z^{*}\mathbf{J}_{-}^{\dagger}} = \mathrm{e}^{- i \, \theta \, (\sin(\varphi) \, \mathbf{J}_1 + \cos(\varphi) \, \mathbf{J}_2)} \, ,
\ee
is the unitary representation on $\mathbb{C}^2$ of a rotation of an angle $\theta$ around the vector $\mathbf{m}= (\sin(\varphi) \, ,  \cos(\varphi), \, 0)$. 
Therefore, we have the following range of variation for the parameters $(\theta, \varphi)$, i.e., we have
\be 
\theta \in \left[ 0, \pi \right[ \;, \quad \varphi \in \left[0 , 2\pi \right[\,,
\ee
and the map given in equation \eqref{eqn: radcliff pure coherent states} defines an immersion of $S^2_{p}$ (the 2-dimensional sphere  without one of the poles) in the Hilbert space associated with a spin-$\frac{1}{2}$ system.
The density matrix $\rho_{z}$ associated with $|z\rangle$ has the form:
\be
\rho_{z}=|z\rangle\langle z|= \left(\begin{matrix} p_{3}(z) & p_{1}(z) - \imath\,p_{2}(z) - \gamma^{*} \\ & \\ p_{1}(z) + \imath\,p_{2}(z) + \gamma & 1 - p_{3}(z)\end{matrix}\right)\,,
\ee
where $\gamma=\frac{1 + i}{2}$, and:
\be
p_{1}(z) - \frac{1}{2}= \frac{\cos(\varphi) \sin(2|z|)}{2}\,,\;\;\;p_{2}(z) - \frac{1}{2}= \frac{\sin(\varphi) \sin(2|z|)}{2}\,,\;\;\;p_{3}(z)=\sin^{2}(|z|)\,.
\ee
The numbers $0\leq p_{1}(z),p_{2}(z),p_{3}(z)\leq 1$ are the probabilities to have, in the Radcliffe coherent state $\rho_{z}$, the spin projection $m=\frac{1}{2}$ on the directions $x,y,z$ respectively.
The particular tomographic probability description of spin-states was discussed in \cite{Manko_Marmo_Ventriglia_Vitale_metric_on_the_space_of_quantum_states_from_relative_entropy_tomographic_reconstruction}.
Using a different decomposition of the matrix $U(g_{\mathbf{n}})$ it is possible to introduce a parameterization of the set of states in terms of a variable belonging to the whole complex plane which is related to the parameters $\left( \theta, \varphi \right)$ by means of the stereographical projection, i.e.
\be
\zeta = \tan \left(\frac{\, \theta}{2} \,\right) \mathrm{e}^{- \imath \varphi} \in \mathbb{C}\,,
\ee
and the new set of coherent states is 
\be
|\zeta \rangle:= \left(\begin{matrix} \frac{\zeta}{\sqrt{1 + |\zeta|^2}}\,  \\ \frac{1}{\sqrt{1+|\zeta|^2}})\end{matrix}\right)\,.
\label{zeta-coherent states}
\ee

The dynamical evolution of the Radcliffe coherent states according to the unitary evolution generated by the Hamiltonian operator:
\be
\mathbf{H}_{R}=\frac{A}{2}\,\left\{\mathbf{J}_{+}\,,\mathbf{J}_{-}\right\} + \frac{B}{2}\,\left[\mathbf{J}_{+}\,,\mathbf{J}_{-}\right]\,,
\ee
is simply:
\be
\mathrm{e}^{\imath t \mathbf{H}_{R}}\,|z\rangle = \mathrm{e}^{\imath \frac{t(A-B)}{2}}\, \left(\begin{matrix} \mathrm{e}^{\imath tB}\,\frac{\sin(|z|)}{|z|}\, z \\ \cos(|z|)\end{matrix}\right)=\mathrm{e}^{\imath \frac{t(A-B)}{2}}\, \left(\begin{matrix} \frac{\sin(|z(t)|)}{|z(t)|}\, z(t) \\ \cos(|z(t)|)\end{matrix}\right)\,,
\ee
with:
\be
z(t)=z\,\mathrm{e}^{\imath tB}\,.
\ee
This represents an example where even though the subset of selected states does not carry a linear structure as a linear subspace of the Hilbert space, the dynamics induced by the quantum evolution is again linear with respect to the linear space structure of the parameter manifold.

\subsection{Bosonic Oscillators}

The formalism of coherent states $|z\rangle$  for the harmonic oscillator \cite{Glauber_Photon_correlations,Sudarshan_equivalence_of_semiclassical_and_quantum_mechanical_descriptions_of_statistical_light_beams}, with annihilation and creation operators, $\mathbf{a}$ and $\mathbf{a}^{\dagger}$ respectively, such that $[\mathbf{a}\,,\mathbf{a}^{\dagger}]=\mathbb{I}$, is based on the definition:
\be
|z\rangle:=\mathrm{e}^{z\mathbf{a} - z^{*}\mathbf{a}^{\dagger}}\,|0\rangle\,.
\ee
The ground (vacuum) state $|0\rangle$ is such that $\mathbf{a}|0\rangle=0 \cdot |0\rangle$, where $\langle 0|0\rangle=1$.
The complex number $z=\frac{1}{\sqrt{2}}(q + \imath p)$ determines the set of parameters $\mathcal{M}$.
The evolution of Bosonic-oscillator coherent states under the action of the oscillator Hamiltonian, i.e.:
\be
|z_{t}\rangle\langle z_{t}|=\mathrm{e}^{-\imath \mathbf{H} t}\,|z\rangle\langle z|\,\mathrm{e}^{\imath\mathbf{H} t}\,,\;\;\;\hbar\equiv 1\,,
\ee
where:
\be
\mathbf{H}=\frac{A}{2}\,\left\{\mathbf{a}^{\dagger}\,,\mathbf{a}\right\} + \frac{B}{2}\,\left[\mathbf{a}^{\dagger}\,,\mathbf{a}\right]\,,
\ee
yields the linear dynamics on the manifold $M$ of the form:
\be
z\mapsto z(t)=z\,\mathrm{e}^{\imath\omega t}\,,
\ee
where $\omega$ is related to the constants $A$ and $B$.

\subsubsection*{The case of $f$-oscillators}

In \cite{Manko_Marmo_Sudarshan_Zaccaria_foscillators_and_nonlinear_coherent_states}, the notion of $f$-oscillator was introduced.
It is a deformation of the oscillator annihilation operator $\mathbf{a}$ of the form:
\be
\mathbf{a}\mapsto \mathbf{A}_{f}:=\mathbf{a}\,f(\mathbf{a}^{\dagger} \, \mathbf{a})=\mathbf{a}\,f(\mathbf{n})\,,
\ee
accompanied by:
\be
\mathbf{a}^{\dagger}\mapsto \mathbf{A}_{f}^{\dagger}:=f(\mathbf{n})^{\dagger} \,\mathbf{a}^{\dagger}\,.
\ee
The properties of $f$-oscillators are discussed in \cite{Dudinec_Manko_Marmo_Zaccaria_tomography_on_foscillators}.
We address the following problem: if we map the manifold $\mathcal{M}=\mathbb{C}$ using  $f$-oscillators, what is the classical-like dynamics we get on $\mathcal{M}$?
We consider the Hamiltonian evolution dictated by:
\be
\mathbf{H}=\frac{\omega}{2}\,\left(\mathbf{A}^{\dagger}_{f}\,\mathbf{A}_{f} + \mathbf{A}_{f}\,\mathbf{A}^{\dagger}_{f}\right)\,,
\label{f-oscillators Hamiltonian}
\ee
for the nonlinear coherent state defined for $f$-oscillators by the eigenvalue equation:
\be
\mathbf{A}_{f}\,|z\rangle_{f}= z\,|z\rangle_{f}\,.
\ee
The nonlinear coherent state can be written as the following superposition of the basis vectors $\left\lbrace | n \rangle \right\rbrace$ \cite{Manko_Marmo_Sudarshan_Zaccaria_foscillators_and_nonlinear_coherent_states}:
\be
|z\rangle_{f} = N_{f}\,\sum_{n=0}^{+\infty}\,\frac{z^{n}}{\sqrt{n!} \,[f(n)]!} \,|n\rangle\,,\qquad  \mathbf{a}^{\dagger}\, \mathbf{a} \, |n\rangle= n \, |n\rangle\,, 
\ee
with 
\be
[f(n)]!=f(n)\cdot f(n-1)\cdots f(1)  \,, \qquad N_f= \left(\, \sum_{n=0}^{\infty} \frac{|\alpha|^{2n}}{n! \, |[f(n)]!|^2}  \,\right)^{-\frac{1}{2}}\,.
\ee
In order for the series defining $|z\rangle_f  $ in the previous equation to be convergent, $f(n) $ can be checked to be of the form 
\be
f(\mathbf{n})=\sqrt{\frac{\lambda\mathbf{n} + \alpha}{\gamma \mathbf{n} + \delta}}.
\ee
The question of how to choose the parameters $\lambda, \alpha, \gamma,\delta$ in such a way that  one has classical-like evolutions on the manifold $\mathcal{M}=\mathbb{C}$, parametrized by $z$, only has the trivial solution $\lambda=\alpha, \gamma=\delta$ so that  
\be
f(\mathbf{n})=\frac{\alpha}{\gamma}\,=\,\mbox{ constant },
\ee
which corresponds to ordinary coherent states with  the classical-like evolution 
\be
z\mapsto z_{t}=z\,\mathrm{e}^{\imath \tilde \omega t}\,,
\ee
which consists in a linear reparameterization of the motion obtained in the previous paragraph.
Any other $f$-function destroys the evolution on the manifold $\mathcal{M}=\mathbb{C}$, since the evolution generated by $\mathbf{H}$ will map f-oscillators into different f-oscillators as the time parameter changes (see \cite{Manko_Marmo_Sudarshan_Zaccaria_foscillators_and_nonlinear_coherent_states} for further details).

\subsection{Fermionic Oscillators}
We have yet another Hilbert space in which we may immerse our manifold, the space of Fermionic states.
The construction of Radcliffe spin-$\frac{1}{2}$ coherent states, in fact, can be applied to construct Fermi-oscillator coherent states.
In the case of Fermi oscillators, the creation and annihilation operators obey the condition:
\be\label{eqn: Fermi cre and ann commuation relation}
\mathbf{c}\,\mathbf{c}^{\dagger} + \mathbf{c}^{\dagger}\,\mathbf{c}=\mathbb{I}\,.
\ee
The two-dimensional realization of the Fermi-oscillator creation and annihilation operators $\mathbf{c}^{\dagger}$ and $\mathbf{c}$ reads:
\be\label{eqn: 2-dim realization of Fermi cre and ann operator}
\mathbf{c}=\left(\begin{matrix} 0 & 0 \\ 1 & 0 \end{matrix}\right)
\,,\;\;\;\;\mathbf{c}^{\dagger}=\left(\begin{matrix} 0 & 1 \\ 0 & 0 \end{matrix}\right)\,.
\ee
The ground state of  Fermi oscillator defined by $\mathbf{c}|0\rangle_{F}=0 \cdot |0 \rangle_F$ gives the state vector:
\be
|0\rangle_{F}=\left(\begin{matrix} 0 \\ 1 \end{matrix}\right)\,.
\ee
The state $|1\rangle_{F}=\mathbf{c}^{\dagger}|0\rangle_{F}$ is given by the vector:
\be
|1\rangle_{F}=\left(\begin{matrix} 1 \\ 0 \end{matrix}\right)\,.
\ee
The number operator for Fermi-oscillator:
\be
\mathbf{n}_{F}=\mathbf{c}^{\dagger}\,\mathbf{c}\,, 
\ee
provides the relation:
\be
\mathbf{n}_{F}|n\rangle_{F}= n|n\rangle_{F}\,,\;\;\;n=0,1\,.
\ee
Equations \eqref{eqn: Fermi cre and ann commuation relation} and \eqref{eqn: 2-dim realization of Fermi cre and ann operator} are Fermi-oscillator counterpart of the Bose-oscillator formulas.
It is a known property of the supersymmetry relating Bose and Fermi-oscillators.
We should stress that already in \cite{Jordan_Der_zusammenhang_der_symmetrischen, Schwinger_quantum_theory_of_angular_momentum} it was observed that the same classical Lie algebra, realized in terms of matrices, could be also realized in terms of the algebra of Bosonic CCR (canonical commutation relations) or in terms of the algebra of Fermionic creation and annihilation operators (ACCR) (also see \cite{Manko_Marmo_Vitale_Zaccaria_A_generalization_of_the_Jordan-Schwinger_map_the_classical_version_and_its_q_deformation}).
On the other hand, the discussed Radcliffe coherent state construction for spin-$\frac{1}{2}$ particles can be applied to the Fermi-oscillator due to the fact that:
\be
\mathbf{c}^{\dagger}=\mathbf{J}_{+}\,,\qquad \mathbf{c}=\mathbf{J}_{-}\,,
\ee
in the $2$-dimensional matrix realizations of $\mathbf{c},\mathbf{c}^{\dagger}$.
Also, the Hilbert space of the Fock state of the Fermi-oscillators coincides with the Hilbert space of the state of the spin-$\frac{1}{2}$ particle:
\be
|0\rangle_{F}=|- 1/2\rangle\,, \qquad |1\rangle_{F}=|1/2\rangle\,.
\ee
In view of this isomorphism, the coherent states of Fermi-oscillators can be constructed formally using the same expressions used for spin-$\frac{1}{2}$ coherent states.
To summarize, the classical-like evolution on $\mathcal{M}$ obtained from the Bose-oscillator, the Fermi-oscillator, and the Radcliffe coherent states are rotations in the plane:
\be
z\mapsto z(t)=z\,\mathrm{e}^{\imath\omega(A,B) t}\,,
\ee
where the frequency $\omega(A,B)$ is determined by the Hamiltonians:
\be
\begin{split}
\mathbf{H}&=\frac{A}{2}\,\left\{\mathbf{a}^{\dagger}\,,\mathbf{a}\right\} + \frac{B}{2}\,\left[\mathbf{a}^{\dagger}\,,\mathbf{a}\right]\,,\\
\mathbf{H}_{R}&=\frac{A}{2}\,\left\{\mathbf{J}_{+}\,,\mathbf{J}_{-}\right\} + \frac{B}{2}\,\left[\mathbf{J}_{+}\,,\mathbf{J}_{-}\right]\,,\\
\mathbf{H}_{F}&=\frac{A}{2}\,\left\{\mathbf{c}^{\dagger}\,,\mathbf{c}\right\} + \frac{B}{2}\,\left[\mathbf{c}^{\dagger}\,,\mathbf{c}\right]\,.
\end{split}
\ee
Also, the Radcliffe coherent states for spin-$\frac{1}{2}$ particles are identical to the coherent states of a single mode Fermi-oscillator. 
Another type of fermionic coherent states was introduced in \cite{Valle_fermionic_coherent_states_in_a_fock_superspace}, but the author used complex Grassmann variables instead of the complex numbers $z$ and $z^{*}$. In a following section we will present an analogous system showing that one can immerse also super-manifolds into a Hilbert space and introduce a suitable dynamics on it starting from a quantum Hamiltonian system.

\subsubsection*{Two-mode Fermi oscillator and Radcliffe coherent states}

For two spin-$\frac{1}{2}$ particles, we can introduce:
\be
\begin{split}
\mathbf{J}_{+}^{1} & =\mathbf{J}_{+}\otimes\mathbb{I}\,,\qquad
\mathbf{J}_{-}^{1}=\mathbf{J}_{-}\otimes\mathbb{I}\,,\\
\mathbf{J}_{+}^{2} & =\mathbb{I}\otimes\mathbf{J}_{+}\,,\qquad
\mathbf{J}_{-}^{2}=\mathbb{I}\otimes\mathbf{J}_{-}\,,
\end{split}
\ee
so that we can define the Radcliffe two-mode coherent states:
\be
|z_{1}\,,z_{2}\rangle=\mathrm{e}^{z_{1}\,\mathbf{J}_{+}^{1} - z_{1}^{*}\,\mathbf{J}_{-}^{1} + z_{2}\,\mathbf{J}_{+}^{2} - z_{2}^{*}\,\mathbf{J}_{-}^{2}}\,|0\,,0\rangle\,,
\ee
where the ground state $|0\,,0\rangle$ is such that $\mathbf{J}_{-}^{1}|0\,,0\rangle=\mathbf{J}_{-}^{2}|0\,,0\rangle=0$.
A direct computation shows that:
\be
| z_1 \, , z_2 \rangle = \left(
\begin{array}{c}
 \frac{\sin |z_1 |}{|z_1 |} \, z_1  \,  \frac{\sin |z_2 |}{|z_2 |} \, z_2 \\
 \frac{\sin |z_1 |}{|z_1 |} \, z_1  \, \cos |z_2 | \\
 \cos |z_1 |  \,  \frac{\sin |z_2 |}{|z_2 |} \, z_2  \\
  \cos |z_1 |  \,  \cos |z_2 |  
\end{array}
\right) \, .
\ee
 For two-modes Fermi-oscillator, the correct representation of the creation and annihilation operators is:
\be
\begin{split}
\mathbf{c}_{1}=\mathbf{c}\otimes\mathbb{I}\,,\;\;\;&\mathbf{c}_{1}^{\dagger}=\mathbf{c}^{\dagger}\otimes\mathbb{I}\,,\\
\mathbf{c}_{2}=\sigma_{3}\otimes\mathbf{c}\,,\;\;\;&\mathbf{c}_{2}^{\dagger}=\sigma_{3}\otimes\mathbf{c}^{\dagger}\,,
\end{split}
\ee
so that we have:
\be
\begin{split}
\mathbf{c}_{1}\mathbf{c}_{2} + &\mathbf{c}_{2}\mathbf{c}_{1} = \mathbf{c}_{1}^{\dagger}\mathbf{c}_{2}^{\dagger} + \mathbf{c}_{1}^{\dagger}\mathbf{c}_{2}^{\dagger} = \mathbf{c}_{1}\mathbf{c}_{2}^{\dagger} + \mathbf{c}_{2}^{\dagger}\mathbf{c}_{1} =0 \,,\\
& \mathbf{c}_{1}\mathbf{c}_{1}^{\dagger}+ \mathbf{c}_{1}^{\dagger}\mathbf{c}_{1} = \mathbf{c}_{2}\mathbf{c}_{2}^{\dagger} + \mathbf{c}_{2}^{\dagger}\mathbf{c}_{2}= \mathbb{I}\,.
\end{split}
\ee
The explicit expression of $\mathbf{U}(z_{1},z_{2})=\mathrm{e}^{z_{1}\mathbf{c}_{1} - z_{1}^{*}\mathbf{c}_{1}^{\dagger} + z_{2}\mathbf{c}_{2} - z_{2}^{*}\mathbf{c}_{2}^{\dagger}}$ reads:
\be
\mathbf{U}(z_{1},z_{2}) = \left(
\begin{array}{cccc}
 \cos \mathbf{z} &  - \frac{\sin \mathbf{z}}{\mathbf{z}} \, z^{\star}_2 & - \frac{\sin \mathbf{z}}{\mathbf{z}} \, z^{\star}_1 & 0 \\
  \frac{\sin \mathbf{z}}{\mathbf{z}} \, z_2  &   \cos \mathbf{z} & 0 &  - \frac{\sin \mathbf{z}}{\mathbf{z}} \, z^{\star}_1 \\
  \frac{\sin \mathbf{z}}{\mathbf{z}} \, z_1 & 0 &\cos \mathbf{z}  &  \frac{\sin \mathbf{z}}{\mathbf{z}} \, z^{\star}_2 \\
0 &  \frac{\sin \mathbf{z}}{\mathbf{z}}\, z_1 & -  \frac{\sin \mathbf{z}}{\mathbf{z}} \, z_2 & \cos \mathbf{z} 
\end{array}
\right) \, ,
\ee
where $ \mathbf{z} := \sqrt{ |z_1|^2 + |z_2|^2} $. Depending on the choice of the fiducial vector, we get quite different behaviours.
For instance, if we choose as fiducial vector the vector $|00\rangle$ such that $\mathbf{c}_{1}|00\rangle=\mathbf{c}_{2}|00\rangle=0\cdot |0\rangle$, we can define:
\be
|z_{1}\,,z_{2}\rangle_{00}:=\mathrm{e}^{z_{1}\mathbf{c}_{1} - z_{1}^{*}\mathbf{c}_{1}^{\dagger} + z_{2}\mathbf{c}_{2} - z_{2}^{*}\mathbf{c}_{2}^{\dagger}}\, |00\rangle
=
\left(
\begin{array}{c}
 0 \\
- \frac{\sin \mathbf{z}}{\mathbf{z}} \, z^{\star}_1 \\
  \frac{\sin \mathbf{z}}{\mathbf{z}} \, z^{\star}_2 \\
  \cos \mathbf{z}
\end{array}
\right) \, .
\ee
It is clear that, even if we started with the two free parameters $z_{1},z_{2}\in\mathbb{C}$, the immersed parameter manifold is actually $\mathcal{M}=\mathbb{C}$.
Let us consider the one parameter group of unitary transformations $\mathbf{U}(t)$ generated by the oscillator Hamiltonian:
\be
\mathbf{H}^{12}_{F}=H_{F}^{1} + H_{F}^{2}=\frac{A_{1}}{2}\,\left\{\mathbf{c}_{1}^{\dagger}\,,\mathbf{c}_{1}\right\} + \frac{B_{1}}{2}\,\left[\mathbf{c}_{1}^{\dagger}\,,\mathbf{c}_{1}\right] + \frac{A_{2}}{2}\,\left\{\mathbf{c}_{2}^{\dagger}\,,\mathbf{c}_{2}\right\} + \frac{B_{2}}{2}\,\left[\mathbf{c}_{2}^{\dagger}\,,\mathbf{c}_{2}\right]\,,
\ee
Since these expressions will be used in the rest of this paragraph, let us define the following set of frequencies depending on the parameters of the Hamiltonian operator:
\be
\begin{split}
\omega_1 & =   \frac{1}{2} (A_1+A_2+B_1+B_2) \, , \qquad
\omega_2 = \frac{1}{2} (A_1+A_2+B_1-B_2) \, , \\
\omega_3 & =  \frac{1}{2} (A_1+A_2-B_1+B_2) \, , \qquad
\omega_4 = \frac{1}{2} (A_1+A_2-B_1-B_2) \, .
\end{split}
\ee
Therefore, when $\mathbf{U}(t)$ acts on $|z_1,z_2 \rangle$ we obtain the dynamics:
\be
z_{1}\mapsto z_{1}(t)=z_{1}\,\mathrm{e}^{\imath \, \omega_2 t} \,, \qquad
z_{2}\mapsto z_{2}(t)=z_{2}\,\mathrm{e}^{\imath \, \omega_3 t}\,,
\ee
of the single-mode Fermi-oscillator. 
What happens when we select $\frac{|10\rangle + |01\rangle}{\sqrt{2}}$ as fiducial state?
A direct computation shows that:
\be
|z_{1}\,,z_{2}\rangle_{01}:=\mathrm{e}^{z_{1}\mathbf{c}_{1} - z_{1}^{*}\mathbf{c}_{1}^{\dagger} + z_{2}\mathbf{c}_{2} - z_{2}^{*}\mathbf{c}_{2}^{\dagger}}\,\left(\frac{|10\rangle + |01\rangle}{\sqrt{2}}\right)=
\frac{1}{\sqrt{2}} \, \left(
\begin{array}{c}
 - \frac{\sin \mathbf{z}}{\mathbf{z}} \, (z^{\star}_1 + z^{\star}_2 )\\
\cos \mathbf{z} \\
\cos \mathbf{z} \\
  \frac{\sin \mathbf{z}}{\mathbf{z}} \, (z_1 - z_2 )
\end{array}
\right) \, ,
\ee
and thus:
\be
\mathbf{U}(t) \,|z_{1}\,,z_{2}\rangle_{01} = \frac{1}{\sqrt{2}} \, 
\left(
\begin{array}{c}
 -\frac{\sin \mathbf{z}}{\mathbf{z}} \, e^{i \, \omega_1 \, t }  \, (z^{\star}_1 + z^{\star}_2 )  \\
e^{i \, \omega_2 \, t } \, \cos \mathbf{z}   \\
 e^{i \, \omega_3 \, t } \, \cos \mathbf{z}    \\
 \frac{\sin \mathbf{z}}{\mathbf{z}} \, e^{i \, \omega_4 \, t }  \, (z_1 - z_2 ) 
\end{array}
\right) \, .
\ee

Consequently, if we want dynamics to map coherent states  into coherent states, it must be $ B_1 = B_2  \equiv - \omega $, so that:

\begin{equation}
\mathbf{U}(t)\,|z_{1}\,,z_{2}\rangle_{01} = \frac{e^{i(A_1+A_2)t}}{\sqrt{2}}  \, 
\left(
\begin{array}{c}
 -\frac{\sin \mathbf{z(t)}}{\mathbf{z(t)}} \,   (z^{\star}_1(t) + z^{\star}_2(t) )  \\
 \cos \mathbf{z(t)}   \\
 \cos \mathbf{z(t)}    \\
 \frac{\sin \mathbf{z(t)}}{\mathbf{z(t)}} \, (z_1(t) - z_2(t) ) 
\end{array}
\right) \, ,
\end{equation}
From this, it follows that the classical-like dynamics  in  terms of the variables is given by:
\begin{equation}
\begin{split}
z_{1} & \mapsto z_{1}(t)=z_{1}\,\mathrm{e}^{\imath \, \omega \, t} \\
z_{2} & \mapsto z_{2}(t)=z_{2}\,\mathrm{e}^{\imath \, \omega \, t}\, , \\
\mathbf{z}& \mapsto  \mathbf{z(t)} =  \sqrt{ |z_1(t)|^2 + |z_2(t)|^2}\,.
\end{split}
\end{equation}
We have thus found that the same dynamics on a classical-like manifold  may be obtained by immersing the manifold  into different quantum models, suggesting that our immersion should not be thought of as a quantization.

\begin{remark}[Probabilities]
Quantum states in the Schr\"{o}dinger picture can be interpreted as probability amplitudes. Therefore, the procedure outlined above allows us to associate probability amplitudes with the points of a manifold. On the other hand, one could immerse the same space directly in the space of probability distributions: in this case the representation of spin-$\frac{1}{2}$ coherent states and Fermi-oscillator coherent states provide rotation of probabilities.
Indeed, in the tomographic probability representation (see \cite{Ibort_Manko_Marmo_Simoni_Ventriglia_A_pedagogical_presentation_of_tomography} for an introduction to the tomographic picture of Quantum Mechanics) the dynamical evolution is expressed in terms of linear equations for probabilities $p_{1}(z,t),p_{2}(z,t),p_{3}(z,t)$.
In vector notation we have:
\be
\frac{\mathrm{d} \vec{p}(t)}{\mathrm{d}t}= \mathbf{L}(t) \vec{p}(t) + \vec{r}(t)\,,
\ee
where:
\be
\vec{p}(t)=\left(\begin{matrix} p_{1}(z,t) \\ p_{2}(z,t) \\ p_{3}(z,t)\end{matrix}\right)\,.
\ee
There, the $(3\times 3)$ matrix $\mathbf{L}(t)$ reads:
\be
\mathbf{L}(t)=\frac{1}{2}\left(\begin{matrix} 0 & -2B & 0  \\ 2B &0 & 0 \\ 0& 0& 0 \end{matrix}\right)\,,
\ee
and the $3$-vector $\vec{r}(t)$ is:
\be
\vec{r}(t)=\frac{1}{2}\left(\begin{matrix} B \\ -B \\0 \end{matrix}\right)\,.
\ee
\end{remark}

\begin{remark}[Cat states for Radcliffe and Fermi-oscillator coherent states]
In \cite{Dodonov_Malkin_Manko_even_and_odd_coherent_states}  even and odd coherent states (also called even and odd cat states) of oscillators were introduced as the superposition:
\be\label{eqn: even cat states boson}
|z_{\pm}\rangle=N_{\pm}\left(|z\rangle \pm |-z\rangle\right)\,,
\ee
where:
\be
\begin{split}
N_{+} & = \left[ \, \sum_{n=0}^{+\infty}\,\frac{z^{2n}}{\sqrt{2n!}} \, \right]^{-\frac{1}{2}} \, , \\
&\\
N_{-} & = \left[ \, \sum_{n=0}^{+\infty}\,\frac{z^{2n + 1}}{\sqrt{(2n + 1)!}}  \, \right]^{-\frac{1}{2}} \, .
\end{split}
\ee
The superposition for two-mode coherent states \cite{Ansari_Manko_photon_statistics_of_multimode_even_and_odd_coherent_lght} is an entangled state.
The evolution of the cat state under the action of the oscillator Hamiltonian provides again the harmonic oscillator dynamics on the parameter $z$:
\be
z\mapsto z_{t}=z\,\mathrm{e}^{\imath\omega t}\,.
\ee
Since we introduced Fermi-oscillator coherent states we introduce even and odd cat states for this oscillator and study their properties in the sense of classical-like dynamics.
Using equation \eqref{eqn: radcliff pure coherent states} we obtain:
\be
|z_{+}\rangle^{R}_{F}=N_{+} \biggl( \, \cos |z| \; |0\rangle + \sin |z| \, \cos \gamma(z) \; |1\rangle \, \biggr) \, ,
\ee
\be
|z_{-}\rangle^{R}_{F}=N_{-} \, \sin |z| \, \sin \gamma(z) \; |1 \,\rangle\,.
\ee
It is then clear that the induced classical-like evolution on $z$ is the same as above.
The density matrix of even cat states is:
\be
\rho_{+}=|z_{+}\rangle^{R}_{F} \langle z_{+}| = |N_{+}|^{2} \,
\left(
\begin{matrix} 
\sin^{2} |z| \, \cos^2 \gamma(z)  &  \frac{1}{2} \, \sin 2|z| \, \cos \gamma(z)  \\ 
\frac{1}{2} \, \sin 2|z| \, \cos \gamma(z)  & \cos^{2} |z|\end{matrix}
\right)\,,
\ee
while the density matrix of odd cat states is:
\be
\rho_{-} = |z_{-}\rangle^{R}_{F} \langle z_{-}| = |N_{-}|^{2} \, \left(
\begin{matrix} 
\sin^{2} |z| \, \sin^{2} \gamma(z)  & 0 \\ 
0 & 0
\end{matrix}\right)\,.
\ee
\end{remark}

\section{Reduction by means of Lagrangian formalism} \label{Sec: lagrangian formalism}

As already mentioned in the introduction, the procedure described in previous examples only works if the immersed submanifold of states is invariant under the unitary evolution generated by a given Hamiltonian operator. 
Therefore,  in order to make the procedure work  for every immersed submanifold, $\mathcal{M}$, one should have a description of dynamics in terms of covariant tensors that may be pulled back on $\mathcal{M}$ by means of the immersion map $\mathfrak{i}$ giving rise to a dynamics on $\mathcal{M}$. 
In this section, we will discuss how  the {Lagrangian formalism} may be used for this purpose. Indeed, if a Lagrangian function $\lag \in \mathscr{F}(\textbf{T}\mathcal{H})$ for equation \eqref{Eq: schrodinger equation} is available,   the pull-back of $\lag$ via the tangent map $T\mathfrak{i}$, i.e., $\tilde{\lag}=(T\mathfrak{i})^*\lag$, can be used as a Lagrangian function on the tangent bundle $\textbf{T}\mathcal{M}$ to derive the corresponding Euler-Lagrange equations.

The first step, therefore, consists in defining a Lagrangian function $\lag$ for the Schr\"{o}dinger equation. 
At this purpose, we recall that, since states in quantum mechanics are equivalence classes of vectors in $\mathcal{H}$, any class describing a ``ray'', the dynamical evolution of quantum systems should be conceived not on the whole Hilbert space, but rather on its complex projective space, that is, the space obtained by the following quotient
\be \label{Eq: Projective space bundle}
\begin{tikzcd}
\mathcal{H}_{0} \arrow[dd, "\pi"] \arrow[dr, "\mathbb{R}^+"] & \\
 & \mathcal{S}(\mathcal{H}) \arrow[ld, "\mathcal{U}(1)"] \\
\mathcal{P}(\mathcal{H}) &
\end{tikzcd}
\ee
where $\mathcal{H}_{0}$ is the Hilbert space without the null vector.
Accordingly, we will look for a Lagrangian function on $\mathbf{T}\mathcal{H}_{0}$  which is invariant (up to a total time derivative) under the action of $\mathbb{R}^+$ and $\mathcal{U}(1)$ associated with the double fibration of equation \eqref{Eq: Projective space bundle}. 
A possibility is given by
\be \label{Eq: Lagrangian projective}
\lag \,=\, \frac{i}{2} \left[\,\frac{(\psi,\,v_\psi)\,-\,(v_\psi,\,\psi)}{(\psi,\,\psi)} \,\right] \,-\,\frac{(\psi,\,{\mathbf{H}}\psi)}{(\psi,\,\psi)} \,\,.
\ee
The {principle of least action} applied to equation \eqref{Eq: Lagrangian projective} gives rise to the following equation (see \cite[pp. 4-5]{Kramer_Saraceno_Geometry_of_the_time_dependent_variational_principle_in_Quantum_Mechanics} for details)
\be \label{Eq: projective Schrodinger equation}
 i\frac{d}{dt} \psi \,=\, \mathbf{H} \psi - \frac{ \left(\,\psi,\,\left[ i\frac{d}{dt} - {\mathbf{H}} \right] \psi \, \right)}{(\psi,\,\psi)} \, \psi \,.
\ee
which does not coincide with equation \eqref{Eq: schrodinger equation}. Let us notice that, since the Lagrangian function is first-order in the velocities, Euler-Lagrange equations can be written as a set of first-order differential equations on $\mathcal{H}_0$, defining therefore a vector field on $\mathcal{H}_0$ \footnote{In general Euler-Lagrange equations could not define a vector field, for instance in presence of constraints. As we will see, in such a case a careful analysis of the constraints has to be performed in order to get a vector field.}. Furthermore, equation \eqref{Eq: projective Schrodinger equation} is invariant with respect to multiplication by a non-zero complex number in the sense that, if $|\psi \rangle$ is a solution with initial datum $|\psi_0\rangle$, also $\lambda(t)|\psi \rangle$ ($\lambda \in \mathbb{C}_0 \, \forall t \in \left[ t_0,t_1 \right]$) is a solution with initial datum $\lambda(0)|\psi_0\rangle$. This implies that the solutions of equation \eqn{Eq: projective Schrodinger equation} project down to the complex   projective space $\mathcal{P}(\mathcal{H})$. 
As a remark, let us notice that, despite the invariance properties of the equations of motion, the Lagrangian \eqref{Eq: Lagrangian projective} cannot be the pullback to $\mathbf{T}\mathcal{H}$ of a continuous Lagrangian function on $\mathbf{T}\mathcal{P}(\mathcal{H})$. Indeed, this would correspond to a global gauge fixing which cannot be achieved on the bundle $T\pi\,:\,\mathbf{T}\mathcal{H}_{0} \, \rightarrow \, \mathbf{T}\mathcal{P}(\mathcal{H})$ because the bundle is not trivializable (for more details see  \cite{Balachandran_Marmo_Skagerstam_Stern_gauge_symmetries_and_fibre_bundles}).

In order to obtain exactly \eqref{Eq: schrodinger equation} one should lift the vector field defined by \eqref{Eq: projective Schrodinger equation} to $\mathcal{H}_0$ in a suitable way. This choice corresponds to the definition of a connection and the associated parallel transport (see \cite{Landi_Marmo_algebraic_differential_calculus_for_gauge_theories} for more details about the definition of a connection from an algebraic point of view, and \cite{Kolar_Michor_Slovak_natural_operations_in_differential_geometry} for a more general approach to connections on fiber bundles).
In this case, if one chooses the following element $\phi(t) $   
\be \label{Eq: Pancharatnam}
\phi(t)\,:=\, e^{i \int_0^t \left(\,\psi,\,\left[ i\frac{d}{dt} - {\mathbf{H}} \right] \psi \, \right) (\psi,\,\psi)^{-1} \,\mathrm{d}\tau } \psi(t)\,,
\ee
as representative of the equivalence class to which $\psi(t)$ belongs, $\phi(t)$  has the same norm for all $t \in \left[ t_0,t_1 \right]$ and one recovers exactly Schr\"odinger equation \eqref{Eq: schrodinger equation} for $\phi$, as it can be proved via a direct, straightforward substitution. 

In summary, we have seen that the Lagrangian function $\lag$ on $\mathbf{T}\mathcal{H}_{0}$ is a Lagrangian function and the related Euler-Lagrange equations can be read as a vector field on the space of the equivalence classes of rays, i.e., the complex projective space $\mathcal{P}(\mathcal{H})$. If one chooses a proper notion of parallel transport, which amounts to fix a normalization and a phase factor, it is possible to recover the usual form of the Schr\"{o}dinger equation \eqref{Eq: schrodinger equation}.  
 
Let us now illustrate how Lagrangian formalism could be used to deal with the problem of inducing a dynamics on submanifolds of pure states through an example coming from section \ref{sec: coherent states and nonlinear embeddings}. 

\begin{example}{\textbf{Radcliffe coherent states}}
Let us consider the immersion of the complex disc $\mathcal{M}$ of radius $\rho < \pi$ via Radcliffe coherent states $\left\lbrace |z \rangle \right\rbrace \subset \mathcal{H}_0=\mathbb{C}^2-\left\lbrace \mathbf{0} \right\rbrace$. The tangent bundle is $\mathbf{T}\mathcal{H}_0\cong \mathcal{H}_0\times \mathbb{C}^2 \cong \mathbb{R}^4 - \left\lbrace \mathbf{0} \right\rbrace\times \mathbb{R}^4$, and we consider as coordinate functions $\mathbf{T}\mathcal{H}_0 \ni (\textbf{x},\textbf{v}_{x})\, \rightarrow \, (x^1,x^2,x^3,x^4,\dot{x}^1,\dot{x}^2,\dot{x}^3,\dot{x}^4)$. Let $\textbf{H}= \frac{B}{2} \left[ \mathbf{J}_+ , \mathbf{J}_- \right]$ be the Hamiltonian operator for the quantum dynamics.
Therefore, the Lagrangian function of equation \eqref{Eq: Lagrangian projective} takes the following form:
\be
\lag = \frac{x^j \Omega_{jk} \dot{x}^k}{x_m x^m} -  \frac{B}{2} \frac{x^j L_{jk} x^k}{x_m x^m}  ,
\ee
where we have used Einstein's summation rule for repeated indices and we have introduced the following matrices:
\be
\Omega = \left(  
\begin{array}{cccc}
0 & 1 & 0 & 0 \\
-1 & 0 & 0 & 0 \\
0 & 0 & 0 & 1 \\
0 & 0 & -1 & 0
\end{array}
\right)\,,
\qquad 
L = \left(  
\begin{array}{cccc}
1 & 0 & 0 & 0 \\
0 & 1 & 0 & 0 \\
0 & 0 & -1 & 0 \\
0 & 0 & 0 & -1
\end{array}
\right)\,.
\ee
In order to write the equations of motion, we consider the Cartan 1-form
\be
\vartheta_{\lag} = \frac{\partial \lag}{\partial \dot{x}^k} dx^k = \frac{x^j\Omega_{jk}\mathrm{d}x^k}{x_m x^m}\,,
\ee
and the Lagrangian 2-form 
\be 
\omega_{\lag} = \mathrm{d}\vartheta_{\lag} \,=\, \frac{1}{x_m x^m} \left( \Omega_{jk}  \mathrm{d}x^j\wedge \mathrm{d}x^k - \frac{2x^l\Omega_{jk}x^j }{x_r x^r} dx^l \wedge dx^k  \right)\,. 
\ee
Both these differential forms are the pullback to $\mathbf{T}\mathcal{H}_0$ of a 1-form and a 2-form, respectively, on $\mathcal{H}_0$ because the Lagrangian function is first-order in the generalized velocities. With an abuse of notation, we will call $\theta_{\lag}$ and $\omega_{\lag}$ these forms and in what follows we will refer to these objects defined directly on the Hilbert space\footnote{These forms on $\mathcal{H}_0$ can be obtained after choosing a global section of the vector bundle $\mathbf{T}\mathcal{H}_0$, for instance the zero section $s_0 : \mathcal{H}_0 \rightarrow \mathbf{T}\mathcal{H}_0$, which permits the identification of $\mathcal{H}_0$ with the submanifold $v_{\psi}=0$ in $\mathbf{T}\mathcal{H}_0$. Therefore one could define the two differential forms $\theta^{0}_{\lag} = (s_0)^*(\theta_{\lag})$ and $\omega^{0}_{\lag} = (s_0)^*(\omega_{\lag})$.}. Therefore, the 2-form $\omega_{\lag} \in \Omega_2(\mathcal{H}_0)$ has a kernel spanned by two vector fields which are the generators of the action of the Abelian group $\mathbb{R}_+ \times U(1)$. These two vector fields read as follows:  
\begin{eqnarray}
& \hat{\Delta} = x^j \frac{\partial }{\partial x^j}\,, \\
& X_3 = x^j \Omega_{j}^{k}\frac{\partial }{\partial x^k}\,,
\end{eqnarray}   
where the indices are raised via the identity matrix $\delta^{jk}$. 
This means that the degenerate 2-form $\omega_{\lag}$ is the pullback of a symplectic form on the complex projective space. 
As previously outlined, the lifting procedure in Eq.\ref{Eq: Pancharatnam} defines a parallel transport on $\mathcal{H}_0$ (see \cite{Kolar_Michor_Slovak_natural_operations_in_differential_geometry} for a detailed exposition of the theory of connections on principal bundles) and the lifted path $\phi(t)=\mathbf{\hat{x}}(t)$ will obey the following equations:
\begin{eqnarray*}
& \hat{x}^j(t) \dot{\hat{x}}_j = 0\,, \\
& \hat{x}^j \Omega_{jk} \dot{\hat{x}}^k = E_{\lag}(\mathbf{\hat{x}})\,,
\end{eqnarray*}
where $E_{\lag}$ is the function on $\mathcal{H}_0$ defined by the expectation value of the Hamiltonian operator $\textbf{H}$, that is, we have
\be
E_{\lag}=\dot{x}^j \frac{\partial \lag}{\partial \dot{x}^j} - \lag = \frac{B}{2} \frac{x^j L_{jk} x^k}{x_m x^m}\,.
\ee
Let us notice once more that, since the Lagrangian function is first-order in the generalized velocities, $E_{\lag}$ is the pull-back of a function on $\mathcal{H}_0$ and, as already remarked, we will call again $E_{\lag}$ such a function on $\mathcal{H}_0$. 
The first equation is imposing the conservation of the norm of the vector along the curve, whereas the second condition is fixing the evolution of the remaining phase factor. These two conditions will permit to obtain a solution of Schr\"{o}dinger equation starting from an equivalence class of solutions of Euler-Lagrange equations of motion.

The vector field tangent to the curve $\mathbf{\hat{x}}(t)$ is the desired lift $\Gamma_{l}$ on $\mathcal{H}_0$ of the  vector field $\Gamma$ on $\mathcal{P}(\mathcal{H})$, and the equations of motion associated with it can be written as
\be
i_{\Gamma_{l}}\omega_{\lag} = -\mathrm{d}E_{\lag}\,, 
\ee
where
\begin{equation*}
\mathrm{d}E_{\lag} = \frac{B}{x_m x^m} \left( x^j L_{jk} dx^k - \frac{x^j L_{jk}x^k}{x_n x^n} x_l dx^l \right)\,.
\end{equation*} 

A straightforward computation shows that the vector field $\Gamma_{l}$ has the following expression
\be
\Gamma_{l} = \frac{B}{2}\left[ \left( x^1\frac{\partial}{\partial x^2} - x^2\frac{\partial}{\partial x^1} -  x^3\frac{\partial}{\partial x^4} + x^4\frac{\partial}{\partial x^3} \right)\right]    \,,
\ee
and its flow on $\mathcal{H}_0$ can be expressed as
\be
\left( 
\begin{array}{c}
z_1(0) \\
z_2(0)
\end{array}
\right)
\rightarrow 
\left( 
\begin{array}{c}
z_1(0) \mathrm{e}^{i\frac{B}{2}t} \\
z_2(0) \mathrm{e}^{-i\frac{B}{2}t}
\end{array}
\right)\,.
\ee

Let us now consider the immersion defined by equation \eqn{eqn: radcliff pure coherent states}. The pullback $\tilde{\lag}:= (Ti)^*\lag$ is the Lagrangian function
\be 
\tilde{\lag}=\dot{\varphi} \sin^2\left( \frac{\theta}{2} \right) - \frac{B}{2}\cos (\theta)\,,
\ee
which is first-order in the generalized velocities, and the corrsponding Cartan 1-form is
\be 
\tilde{\vartheta}_{\lag} = \sin^2\left( \frac{\theta}{2} \right)d\varphi\,.
\ee
The Lagrangian 2-form is $\tilde{\omega}_{\lag} = \mathrm{d}\tilde{\vartheta}_{\lag} = \frac{1}{2}\sin (\theta) d\theta \wedge d\varphi $ and one can notice that is the pullback of a symplectic 2-form on the manifold $\mathcal{M}$, because the Lagrangian function is first-order in the generalized velocities. Eventually, one can compute also $\mathrm{d}\tilde{E}_{\lag} = -\frac{B}{2} \sin(\theta) d\theta$ and, recalling the previous remarks concerning first-order Lagrangian functions, the equations of motion can be written as follows:
\be
i_{\tilde{\Gamma_{l}}} \tilde{\omega_{\lag}} = -\mathrm{d}\tilde{E}_{\lag}\,,
\ee
where $\tilde{\Gamma_{l}} = B\frac{\partial}{\partial \varphi}$. 
We can see that in this case the solutions of the equations of motion project down to rotations on the disc preserving the radius, which is thus a constant of the motion. Since we have started with a submanifold $\Sigma : = i(\mathcal{M}) \subset \mathcal{H}$ which was invariant under the unitary evolution generated by $\mathbf{H}$, the resulting dynamical evolution coincides with the dynamical evolution obtained in section \ref{sec: coherent states and nonlinear embeddings}    by means of the contravariant-like procedure. This is, actually, a general result which will be shown in the rest of this section.
\end{example}

Now, we have two groups of objects.
On  one hand, on $\textbf{T}\mathcal{H}_0$ we can define
\be \label{Eq: diagram pull-back Lagrangian}
\begin{tikzcd}
\lag \in \mathscr{F}(\textbf{T}\mathcal{H}_0)  \arrow[d] \arrow[r] & E_\lag \,=\, \dot{\psi}^j \dfrac{\partial \lag}{\partial \dot{\psi}^j}-\lag \in \mathscr{F}(\textbf{T}\mathcal{H}_0) \\
\theta_{\lag} \,=\, \dfrac{\partial \lag}{\partial \dot{\psi}^j} \, \mathrm{d}\psi^j  \in \Omega^1(\textbf{T}\mathcal{H}_0) \arrow[d] \\
\omega_\lag = \mathrm{d}\theta_\lag \in \Omega^2(\textbf{T}\mathcal{H}_0) ,
\end{tikzcd}
\ee
on the other hand, on $\textbf{T}\mathcal{M}$ we have
\be \label{Eq: diagram pull-back Lagrangian}
\begin{tikzcd}
(T\mathfrak{i})^* \lag \,=:\, \tilde{\lag} \in \mathscr{F}(\textbf{T}\mathcal{M}) \arrow[d] \arrow[r] & \tilde{E}_\lag \,=\, \dot{\mathbf{m}}^j \dfrac{\partial \tilde{\lag}}{\partial \dot{\mathbf{m}}^j}-\tilde{\lag} \in \mathscr{F}(\textbf{T}\mathcal{M})\\
\tilde{\theta}_\lag := \dfrac{\partial \tilde{\lag}}{\partial \dot{\mathbf{m}}^j}\, \mathrm{d}\mathbf{m}^j \in \Omega^1(\textbf{T}\mathcal{M}) \arrow[d]\\
\tilde{\mathrm{d}} \tilde{\theta}_\lag \,=:\, \tilde{\omega}_\lag \in \Omega^2(\textbf{T}\mathcal{M}) .
\end{tikzcd}
\ee
Let us remark that the index $j$ is labelling coordinate functions on the Hilbert space $\mathcal{H}$, which could be infinite dimensional. 

The following set of equalities is always valid:
\be
\begin{split}
\tilde{\theta}_{\lag} &= (T\mathfrak{i})^*\theta_{\lag},\\
\tilde{\omega}_{\lag} &= (T\mathfrak{i})^*\omega_{\lag}, \\
\tilde{E}_{\lag} &= (T\mathfrak{i})^*E_{\lag}\,.
\end{split}
\ee
In particular, the first equality is a consequence of the fact that $S$, the soldering form on $\textbf{T}\mathcal{H}_0$, and $\tilde{S}$, the soldering form on $\textbf{T}\mathcal{M}$, are $1-1$ tensors that are $T\mathfrak{i}$-related, namely $(T\mathfrak{i})^* S \cdot \mathrm{d}f \,=\, \tilde{S}(T\mathfrak{i})^* \mathrm{d}f$ \cite{Marmo_Neinhuis_operators_in_Classical_Dynamics}. The second equality expresses the commutativity of exterior differential with respect to pullback operations.
The last equality comes from the fact that the two vector fields $\Delta$ and $\tilde{\Delta}$ are $Ti$-related, these two vector fields defining the linear structures of the fibres of $\textbf{T}\mathcal{H}_0$ and $\textbf{T}\mathcal{M}$, respectively. 

As already noticed in the example above, the forms $\theta_{\lag}, \,\omega_{\lag}$ and the function $E_{\lag}$ are the pull-back to $\mathbf{T} \mathcal{H}_0$ of objects on $\mathcal{H}_0$ because the Lagrangian function is first-order in the generalized velocities, and analogous results are valid for the covariant objects on $\textbf{T}\mathcal{M}$. With an abuse of notation we will preserve the same names for the covariant objects on $\mathcal{H}_0$ and $\mathcal{M}$ (for instance $\omega_{\lag}$ will belong to $\Omega^2(\mathcal{H}_0)$ and $\tilde{\omega}_{\lag}$ to the space $\Omega^2(\mathcal{M})$.)   
 
The first group of objects would describe the dynamics on $\textbf{T}\mathcal{H}_0$. However, according to the above remark, the dynamical evolution can be actually written in terms of objects on $\mathcal{H}_0$ and we can write the equations of motion in terms of the vector field $\Gamma_{l}$ on $\mathcal{H}_0$ obeying
\be \label{Eq: Intrinsic Euler-Lagrange in Hamiltonian form}
i_{\Gamma_{l}} \omega_\lag \,=\, - \mathrm{d}E_\lag\,.
\ee
Analogously, starting with the objects belonging to the second group, we get the dynamical evolution on $\mathcal{M}$, which is associated with a vector field, say $\tilde{\Gamma_{l}}$, obeying
\be  \label{Eq: intrinsic Euler-Lagrange in Hamiltonian form approximated}
i_{\tilde{\Gamma_{l}}} \tilde{\omega}_\lag \,=\, - \tilde{\mathrm{d}}\tilde{E}_\lag\,.
\ee
 
We will widely use the previous results all over the rest of this section. In particular we will consider directly the pull-back of the forms and not of the Lagrangian functions in the subsequent examples.

Before moving on, let us add some considerations. Indeed, the result of the whole procedure highly depends on the specific immersion we have chosen. In particular, the following aspects must be analysed:
\begin{itemize}
\item The first, obvious, condition that the immersion must satisfy is that it should give rise to a Lagrangian on the parameter-space which is admissible, namely, starting with a Lagrangian admitting strong minima (extrema) we would arrive at a Lagrangian admitting strong minima (extrema).
\item Even if one starts with an even-dimensional Hilbert space and with a Lagrangian giving rise to a two form on $\mathcal{H}_0$ which is the pullback of a symplectic structure on the projective Hilbert space $\mathcal{P}\left( \mathcal{H}\right)$, this property of $\lag$ may be not preserved under pull-back. Namely, the parameter-manifold immersed into the Hilbert space may give rise to a submanifold of $\mathcal{H}_0$ which is isotropic (or Lagrangian) for the symplectic form. This means that the symplectic form vanishes on such a submanifold and, thus, its pull-back to the parameter-manifold is $0$. When this is the case, there is no dymanics on the parameter-manifold. Indeed equations of motion on $\mathcal{M}$ reduce to:
\be
0 \,=\, \tilde{\mathrm{d}}\tilde{E}_{\lag} \,,
\ee
that give an entire subset of $\mathcal{M}$ where one can take initial data that will not evolve in time. This submanifold is exactly the pull-back of the critical submanifold for $E_\lag$. Therefore, by means of immersion into an isotropic (or Lagrangian) submanifold for the "pulled-back" Lagrangian two form one simply recovers the best approximation of the ground state of the Hamiltonian inside the submanifold of states considered (this is what is known as the Variational Method in text-books on Quantum Mechanics). With this in mind, the immersion procedure outlined should be understood in the spirit of the usual Variational Method used to find approximate eigenvalues and eigenvectors of some Hamiltonian by means of submanifolds of selected states which are not invariant under the considered Hamiltonian operator. This covariant-like procedure should be, in fact, understood as a dynamical generalization of the usual Variational Method which provide approximate dynamics "constrained" to subsets of trial states. It should be clear that the specific subset of trial states is in principle completely arbitrary and without additional requirements, for instance providing a resolution of the identity. If in the specific examples considered we will deal with classes of trial states with particular properties, this must be seen only as a computational "facility" rather than as some conceptual constraint.

An intermediate situation is also possible. The parameter-manifold, immersed into the Hilbert space, may give rise to a submanifold which intersects an insotropic or a Lagrangian submanifold for the symplectic form. This means that the symplectic form restricted to such a submanifold turns out to be degenerate and, thus, also its pull-back to $\mathcal{M}$ must be degenerate. Therefore, some components of $\tilde{\Gamma}$ will be in the kernel of $\tilde{\omega}_\lag$ and some equations will not be dynamical equations but only constraint equations.   
\end{itemize}

Having at our disposal a Lagrangian for \eqref{Eq: schrodinger equation} we are now ready to apply the procedure described in the beginning of this section to the examples we addressed in section \eqn{Sec: squeezed correlated states}.

\subsection{Free particle}

As in equation \eqn{Subsec: free particle}, let us consider the free Hamiltonian of equation \eqref{Eq: free hamiltonian} and the submanifold of states given by the immersion define by equation \eqref{Eq: immersion squeezed correlated states}.
Following the construction described in the introduction of the present section, a direct computation gives the following results for the pull-back of the Lagrangian two form and the Lagrangian energy function:
\be \label{Eq: lagrangian two-form squeezed states}
\tilde{\omega}_\lag \,=\, -\left(\frac{b_R^2}{2 a_R^3} + \frac{1}{4 a_R^2} \right) \, \mathrm{d}a_R \wedge \mathrm{d}a_I \,+\, \frac{b_R}{2 a_R^2} \, \mathrm{d}b_R \wedge \mathrm{d}a_I \,+\, \frac{b_R}{2 a_R^2} \, \mathrm{d}a_R \wedge \mathrm{d}b_I \,-\, \frac{1}{2 a_R} \, \mathrm{d}b_R \wedge \mathrm{d}b_I \,\,,
\ee
\be \label{Eq: lagrangian energy squeezed states}
\tilde{E}_\lag \,=\, \frac{a_R}{2} \,+\, \frac{a_I^2}{2 a_R} \,+\, \frac{(a_I b_R - a_R b_I)^2}{2 a_R^2} \,\,.
\ee
Consequently:
\be \label{Eq: differential Lagrangian energy squeezed states}
\begin{split}
\mathrm{d}\tilde{E}_\lag \,=\, &\left(\, \frac{1}{2} - \frac{a_I^2}{2 a_R^2} - \frac{a_I^2 b_R^2}{a_R^3} + \frac{b_R b_I a_I}{a_R^2} \,\right) \mathrm{d}a_R \,+\, \left(\, \frac{a_I}{a_R} + \frac{a_I b_R^2}{a_R^2} - \frac{b_R b_I}{a_R} \,\right) \mathrm{d}a_I \,+ \\
& + \left(\, \frac{b_R a_I^2}{a_R^2} - \frac{b_I a_I}{a_R} \,\right) \mathrm{d}b_R \,+\, \left(\, b_I - \frac{b_R a_I}{a_R} \,\right) \mathrm{d}b_I \,\,.
\end{split}
\ee
Equations of motion given by \eqref{Eq: lagrangian two-form squeezed states} and \eqref{Eq: differential Lagrangian energy squeezed states}:
\be
i_{\tilde{\Gamma}} \tilde{\omega}_\lag \,=\, -\mathrm{d}E_\lag \,,
\ee
provide the following dynamics, $\tilde{\Gamma}$:
\be \label{Eq: approximated dynamics free particle squeezed correlated states}
\begin{split}
\tilde{\Gamma}^{a_R} &\,=\, 4a_R a_I \\ 
\tilde{\Gamma}^{a_I} &\,=\, 2(a_I^2 - a_R^2) \\
\tilde{\Gamma}^{b_R} &\,=\, 2(a_I b_R + b_I a_R) \\ 
\tilde{\Gamma}^{b_I} &\,=\, 2(a_I b_I - a_R b_R) \\
\frac{\partial}{\partial c_R} &,\, \frac{\partial}{\partial c_I} \, \in \, ker\tilde{\omega}_\lag \,\,.
\end{split}
\ee
Of course, the vector fields $\frac{\partial}{\partial c_R}$ and $\frac{\partial}{\partial c_I}$ are in the kernel of $\tilde{\omega}_\lag$ because they are the generators of dilations and multiplication by a phase factor and $\omega_{\lag}$ is the pullback of a form on $\mathcal{P}\left( \mathcal{H} \right)$.

Note that, apart for the latter aspect, which is due to the fact that here we are dealing with a dynamics on $\mathcal{P}(\mathcal{H})$ and not on $\mathcal{S}(\mathcal{H})$, this is exactly the same nonlinear dynamics we found in \eqn{Subsec: free particle}. However, we can lift the vector field $\tilde{\Gamma}$ to a vector field $\Gamma^{\uparrow}$ on $\mathcal{M}$, in such a way that the following parallel transport condition is satisfied:
\begin{equation}
\langle \psi\left( a(t),b(t),c(t) ; x \right) | \left( i \frac{d }{d t} - \mathbf{H} \right) \psi \left( a(t),b(t),c(t) ; x \right) \rangle = 0 \,,
\label{parallel transport M}
\end{equation}  
where $\psi(a(t),b(t),c(t) ; x)$ is any path satisfying \eqref{Eq: approximated dynamics free particle squeezed correlated states}. Such a choice permits to recover also the remaining two equations in \eqref{eqs:restriction free particle}. Since the choice of normalization factors is an arbitrariness which we have shown how to fix, but it doesn't affect the main features of the evolution map, in the rest of this section we will consider the dynamics on the quotient manifold, namely $\hat{\mathcal{M}} \simeq \mathbb{R}_+\times \mathbb{R}^3$.  

One can easily find also the Darboux coordinates for this symplectic dynamical system. Indeed, one can introduce the following set of variables:
\begin{eqnarray}
& q_1 = \frac{b_R}{2a_R} \,, \\
& p_1 = \frac{a_R b_I- a_I b_R}{a_R}\,, \\
& q_2 = \frac{1}{2\sqrt{a_R}}\,, \\
& p_2 = - \frac{a_I}{\sqrt{a_R}}\,, 
\end{eqnarray}
and the 2-form $\tilde{\omega}_{\lag}$ can be written as follows
$$
\tilde{\omega}_{\lag} = \mathrm{d}p_1 \wedge \mathrm{d}q_1 + \mathrm{d}p_2 \wedge \mathrm{d}q_2\,.
$$

The variable $q_1$ is the expectation value of the position operator $\mathbf{x}$ on the Gaussian state $\psi(a,b,c;x)$; $p_1$ is the expectation value of the momentum operator $\mathbf{p}$; $q_2= \Delta x$ is the variance of the operator $\mathbf{x}$, i.e. $(\Delta x)^2 = \langle x^2\rangle - \langle x \rangle^2$, whereas $p_2^2 = \frac{4 (\Delta x)^2(\Delta p)^2 - 1}{4 (\Delta x)^2}$.   

The equations of motion in this new variable become:
\begin{eqnarray}
& \dot{q_1} = p_1\,, \\
& \dot{p_1} = 0\,, \\
& \dot{q_2} = p_2\,, \\
& \dot{p_2} = \frac{1}{4q_2^3}\,,
\end{eqnarray}
and we can notice that the first two equations express Ehrenfest's theorem for quadratic Hamiltonians, whereas the other two equations derive from a Calogero-type potential. There are two independent constants of the motion in involution, which are $p_1$ and $H_2 = p_2^2 + \frac{1}{4q_2^2}$, proving that the system is completely integrable.

\subsection{Harmonic oscillator}
Let us now consider the Hamiltonian of harmonic oscillator \eqref{Eq: hamiltonian harmonic oscillator} and, again, \eqref{Eq: immersion squeezed correlated states} as submanifold of states. 
Of course, the Lagrangian two-form depends only on the chosen immersion. Thus, $\tilde{\omega}_\lag$ turns to be the same as in the previous example:
\be  
\tilde{\omega}_\lag \,=\, -\left(\frac{b_R^2}{2 a_R^3} + \frac{1}{4 a_R^2} \right) \, \mathrm{d}a_R \wedge \mathrm{d}a_I \,+\, \frac{b_R}{2 a_R^2} \, \mathrm{d}b_R \wedge \mathrm{d}a_I \,+\, \frac{b_R}{2 a_R^2} \, \mathrm{d}a_R \wedge \mathrm{d}b_I \,-\, \frac{1}{2 a_R} \, \mathrm{d}b_R \wedge \mathrm{d}b_I \, . \label{Eq: Lagrangian2form}
\ee
On the other side, the pull-back of the mean value of the Hamiltonian operator to the parameters manifold gives:
\be
\tilde{E}_\lag \,=\, \frac{a_R}{2} \,+\, \frac{a_I^2}{2 a_R} \,+\, \frac{(a_I b_R - a_R b_I)^2}{2 a_R^2} + \frac{\omega^2}{8 a_R} + \frac{\omega^2 b_R^2}{8 a_R^2} \,,
\ee
whose differential is:
\be
\begin{split}
\mathrm{d}\tilde{E}_\lag \,=\, &\left(\, \frac{1}{2} - \frac{a_I^2}{2 a_R^2} - \frac{a_I^2 b_R^2}{a_R^3} + \frac{b_R b_I a_I}{a_R^2} - \frac{\omega^2}{8 a_R^2} - \frac{\omega b_R^2}{4 a_R^3} \,\right) \mathrm{d}a_R \,+\, \left(\, \frac{a_I}{a_R} + \frac{a_I b_R^2}{a_R^2} - \frac{b_R b_I}{a_R} \,\right) \mathrm{d}a_I \,+ \\ 
+ &\left(\, \frac{b_R a_I^2}{a_R^2} - \frac{b_I a_I}{a_R} + \frac{\omega b_R}{4 a^2_R} \,\right) \mathrm{d}b_R \,+\, \left(\, b_I - \frac{b_R a_I}{a_R} \,\right) \mathrm{d}b_I \,\,.
\end{split}
\ee
Once again the 2-form is degenerate, the kernel being the same as in previous subsection. In this case, the vector field on $\hat{\mathcal{M}}$ generating the dynamics is:

\begin{eqnarray}
& \tilde{\Gamma}^{a_R} &\,=\, 4a_R a_I\,, \label{covariant-like harmonic oscillator_1} \\ 
& \tilde{\Gamma}^{a_I} &\,=\, 2(a_I^2 - a_R^2) + \frac{\omega^2}{2}\,, \label{covariant-like harmonic oscillator_2}\\
& \tilde{\Gamma}^{b_R} &\,=\, 2(a_I b_R + b_I a_R)\,, \label{covariant-like harmonic oscillator_3}\\ 
& \tilde{\Gamma}^{b_I} &\,=\, 2(a_I b_I - a_R b_R). \label{covariant-like harmonic oscillator_4}
\end{eqnarray}

Again we can stress the fact that apart from the choice of an element in the kernel of the Lagrangian 2-form, this is exactly the same nonlinear dynamics we obtained in \eqn{Subsec: harmonic oscillator}. This dynamics can be recovered if the lifted vector field satisfies \eqref{parallel transport M}, as explained in previous subsection.    

Since the 2-form is the same as the free particle one, the same Darboux coordinates can be introduced in this second case, and one can interpret the results as in the previous one. Indeed, the equations of motion can be expressed as
\begin{eqnarray}
& \dot{q_1} = p_1\,, \\
& \dot{p_1} = -\omega^2 q_1\,, \\
& \dot{q_2} = p_2\,, \\
& \dot{p_2} = -\omega^2 q_2 + \frac{1}{4q_2^3}\,,
\end{eqnarray}
where the first two equations are related to Ehrenfest's theorem, while the remaining ones can be derived from the superposition of a harmonic and a centrifugal (or Calogero-like) potential. This system is completely integrable and two independent constants of motion in involution are $H_1= p_1^2+\omega^2 q_1^2$ and $H_2=p_2^2+ \omega^2 q_2^2 + \frac{1}{4q_2^2}$.

As a final remark let us notice that if one introduce two new complex variables as follows:
\begin{eqnarray}
&z = \frac{b_I - i b_R}{\omega + 2(a_R + i a_I)} \\
&u = \frac{\omega - 2(a_R + i a_I)}{\omega + 2(a_R + i a_I)}\,.
\end{eqnarray} 
It is easy to check that these new variables obey the following equations of motion:
\begin{eqnarray}
&\dot{z} = -i\omega z \\
&\dot{u} = -2i \omega u \,,
\end{eqnarray}
and consequently we have the solutions:
\begin{eqnarray}
& z(t) = z_0 \mathrm{e}^{-i\omega t} \\
& u(t) = u_0 \mathrm{e}^{-2i\omega t} \,.
\end{eqnarray}
These solutions coincides with the solutions obtained in \cite{almalki_kisil-geometric_dynamics_of_a_harmonic_oscillator} if one performs the following substitutions:
\begin{eqnarray}
& E = 2 a_R \qquad x_2 = 2 a_I \\
& Ex_1 = 2  b_R \qquad x_3 = 2 b_I\,.
\end{eqnarray}
Therefore, using this replacement we could also read the dynamics \eqref{covariant-like harmonic oscillator_1}-\eqref{covariant-like harmonic oscillator_4} as a dynamics on a homogeneous space under the action of the step 3 nilpotent Lie group $\mathbb{G}$, according to the notation in \cite{almalki_kisil-geometric_dynamics_of_a_harmonic_oscillator}. However we need to use also the parameter $E$ to reach such a comparison. A more detailed comparison will be addressed somewhere else. 

\paragraph{Anharmonic oscillator}

In the previous paragraphs we have focused on quantum systems for which the selected submanifold was invariant under the quantum evolution map. Consequently, also the Lagrangian formulation provided equations of motion the solution of which coincided with the solution of the Schr\"{o}dinger equation of the initial quantum system. In this paragraph, instead, we are going to present an example in which the submanifold of states in the Hilbert space is not invariant under the quantum evolution. However, the Lagrangian description we have outlined in this section can be applied and it produces a set of equations of motion on a finite dimensional manifold. Such dynamical system provides an approximation of the quantum dynamics, which can be computed using all the methods associated with finite systems of ODEs, in a way which is similar to the variational method used for the eigenvalue problem associated with a given Hamiltonian operator. 
 
In this turn, let us take into account the Schr\"{o}dinger equation associated with the Hamiltonian of the anharmonic oscillator:
\be
i\frac{d}{dt} \psi \,=\, -\frac{1}{2}( \Delta - \omega^2 x^2 - \lambda x^4 ) \, \psi\,, \;\;\;\;\; \psi \in \mathfrak{L}^2(\mathbb{R}) \,,
\label{Schrodinger equation anharmonic oscillator}
\ee
and the submanifold of states \eqref{Eq: immersion squeezed correlated states} as immersion of the parameter space into $\mathfrak{L}^2(\mathbb{R})$. Concerning quantum evolution, it can be shown that for generic values of the coupling constant $\lambda$ the submanifold of Gaussian states is not invariant under the evolution generated by \eqref{Schrodinger equation anharmonic oscillator} (see for instance \cite{Krivoshlykov_Manko_Sissakian_coherent_state_evolution_for_the_quantum_anharmonic_oscillator}, where some details about the evolution generated by the Hamiltonian of the anharmonic oscillator restricted to the submanifold of harmonic coherent states are investigated, and references therein). However, the Lagrangian associated with \eqref{Schrodinger equation anharmonic oscillator} can be pulled-back to the submanifold of Gaussian states and one can obtain an approximate dynamical system. Indeed, by keeping track of the results obtained above, the Lagrangian two-form~\eqref{Eq: Lagrangian2form} remains the same, while the pulled-back energy to the parameter manifold yields,
\be
\tilde{E}_\lag \,=\, \frac{a_R}{2} \,+\, \frac{a_I^2}{2 a_R} \,+\, \frac{(a_I b_R - a_R b_I)^2}{2 a_R^2} + \frac{\omega^2}{8 a_R} + \frac{\omega^2 b_R^2}{8 a_R^2} +
\frac{3 \lambda}{32 a_R^2} + \frac{3 b_R^2 \lambda}{16 a_R^3} + \frac{b_R^4 \lambda}{32 a_R^4} \, .
\ee
Its differential is,
\be
\begin{split}
\mathrm{d}\tilde{E}_\lag \,=\, &\left(\, \frac{1}{2} - \frac{a_I^2}{2 a_R^2} - \frac{a_I^2 b_R^2}{a_R^3} + \frac{b_R b_I a_I}{a_R^2} - \frac{\omega^2}{8 a_R^2} - \frac{\omega b_R^2}{4 a_R^3}  
-\frac{b_R^4 \lambda }{8 a_R^5}-\frac{9 b_R^2 \lambda }{16 a_R^4}-\frac{3 \lambda }{16 a_R^3}
\,\right) \mathrm{d}a_R \,+  \\ 
& \left(\, \frac{b_R a_I^2}{a_R^2} - \frac{b_I a_I}{a_R} + \frac{\omega b_R}{4 a^2_R} +\frac{b_R \omega ^2}{4 a_R^2}
+ \frac{b_R^3 \lambda }{8 a_R^4}+\frac{3 b_R \lambda }{8 a_R^3}
 \,\right) \mathrm{d}b_R \,+ \\
 & \, \left(\, \frac{a_I}{a_R} + \frac{a_I b_R^2}{a_R^2} - \frac{b_R b_I}{a_R} \,\right) \mathrm{d}a_I \,+ \left(\, b_I - \frac{b_R a_I}{a_R} \,\right) \mathrm{d}b_I \,\,.
\end{split}
\ee
and the associated dynamics on the parameter manifold $\hat{\mathcal{M}}$, is:
\be
\begin{split}
\tilde{\Gamma}^{a_R} &\,=\, 4a_R a_I\,, \\  
\tilde{\Gamma}^{a_I} &\,=\, 2(a_I^2 - a_R^2) + \frac{\omega^2}{2} + \frac{3 \lambda  \left( a_R+ b_R^2\right)}{ 4 a_R^2}\,,\\
\tilde{\Gamma}^{b_R} &\,=\, 2(a_I b_R + b_I a_R)\,, \\ 
\tilde{\Gamma}^{b_I} &\,=\, 2(a_I b_I - a_R b_R) + \frac{ b_R^3 \lambda }{2 a_R^3}\,.
\end{split}
\ee

The replacement of an infinite dimensional dynamical system with a finite dimensional one could lead to faster approximation techniques, similar to the ones introduced for the stationary variational method. However, an extensive comparison between solutions of the equations of the motion on the submanifold of Gaussian states and their quantum evolution is beyond the scope of this paper and will be addressed elsewhere.

\subsection{Grassmann variables: the definition of a Lagrangian super-function}
In this last example we are going to consider a different form of the Radcliffe coherent states, where we replace complex number $z$ with complex Grassmann number $\xi$. This map, therefore, defines an immersion of a supermanifold $\mathcal{S}$ into the Hilbert space $\mathcal{H} = \mathbb{C}^2$, and the Lagrangian procedure previously illustrated can be employed, provided that all the consequences on differential calculus due to the non commutativity of Grassmann variables are taken into account. In particular we refer to \cite{Ibort_Landi_Marmo_Solano_on_the_inverse_problem_of_lagrangian_supermechanics,Ibort_Solano_geometrical_foundations_of_lagrangian_supermechanics_and_supersymmetry} and reference therein for more details about Lagrangian formulation on super-manifolds.

\noindent Let $\xi$ be a complex Grassmann variable and let $ |\xi \rangle $ be the state defined as follows:
\be
|\xi \rangle = \mathrm{e}^{\eta\xi \mathbf{J}_+ - \xi^*\eta^* \mathbf{J}_-}|-\rangle = \left(\begin{matrix} \eta\xi \,  \\ 1-\frac{\rho^2}{2} \end{matrix}\right)\,,
\label{grassman coherent states}
\ee
where $\eta$ is a constant Grassmann number such that the product $\eta\xi$ behaves like a commutative number, and $\rho^2 = \eta^*\eta \,\xi^*\xi $. 

\noindent As in the previous cases, this map defines an immersion of the supermanifold $\mathcal{S}$ with a single complex Grassmann variable, in $\mathbb{C}^2$. Via this immersion we obtain the Lagrangian function $\tilde{\lag}$ as the pullback of the Lagrangian function $\lag$ in Eq.\eqref{Eq: Lagrangian projective}: 
\be
\tilde{\lag} = \eta^*\eta \left( \xi_2 \dot{\xi}_1 - \xi_1\dot{\xi}_2 \right) - \frac{A}{2}\rho^2\,,
\ee
where $\xi = \xi_1+i\xi_2$, the Hamiltonian operator is $\mathbf{H}_R$ and we have neglected an inessential constant in the definition of $\tilde{E}_{\lag}$. According to the geometrical description of Lagrangian supermechanics given in \cite{Ibort_Solano_geometrical_foundations_of_lagrangian_supermechanics_and_supersymmetry}, we can define, \textit{mutatis mutandis}, the equations of the motion associated with $\lag$ as follows:
\be 
i_{\tilde{\Gamma}}\tilde{\omega}_{\lag}= -\mathrm{d}\tilde{E}_{\lag}\,,
\ee
where:
\begin{eqnarray*}
& \tilde{\omega}_{\lag}= \mathrm{i}\eta^*\eta \left( \mathrm{d}\xi_1\wedge \mathrm{d}\xi_1 + \mathrm{d}\xi_2 \wedge \mathrm{d}\xi_2 \right)\,, \\
& \mathrm{d}\tilde{E}_{\lag} = \mathrm{i} A \eta^*\eta \left( \xi_1\mathrm{d}\xi_2 - \xi_2 \mathrm{d}\xi_1 \right)\,, \\
&\tilde{\Gamma} = \frac{A}{2} \left( \xi_2 \frac{\partial}{\partial \xi_1} - \xi_1 \frac{\partial}{\partial \xi_2} \right)\,.
\end{eqnarray*} 

Therefore, these equations of the motion define a rotation in the space $\left( \xi_1,\,\xi_2 \right)$. It is worth noticing that in this case, the bilinear operation which can be defined on the space of functions $\mathcal{F}(\mathcal{S})$ via the two-form $\tilde{\omega}_{\lag}$, i.e.
\be 
\left\lbrace f , g \right\rbrace_+ = \tilde{\omega}_{\lag}\left( X_f, X_g \right)\,, 
\ee
where $X_f,X_g$ are the Hamiltonian vector fields associated with $f,g$ via the formula:
\be
i_{X_f}\tilde{\omega}_{\lag}= df,
\ee
is symmetric. In particular we have that:
\begin{eqnarray*}
& \left\lbrace \xi_1 , \xi_1 \right\rbrace_+ = \left\lbrace \xi_2 , \xi_2 \right\rbrace_+ = 1 \,,\\
& \left\lbrace \xi_1 , \xi_2 \right\rbrace_+ = 0\,,
\end{eqnarray*}
which reproduce the usual anticommutation relations for Fermionic operators. Therefore, the same Lagrangian function is, actually, capable of providing both canonical commutation and anticommutation relations. This has to be related to the fact that different representations of the same algebra can be obtained by using complex commutative variables or complex Grassmann variables, as shown in \cite{Manko_Marmo_Vitale_Zaccaria_A_generalization_of_the_Jordan-Schwinger_map_the_classical_version_and_its_q_deformation}.  

\noindent A completely analogous computation could be performed using the Fermi operators $\mathbf{c},\mathbf{c}^{\dagger}$ acting on the Hilbert space $\mathcal{H}_F$ of a single Fermionic particle. This procedure would lead to the coherent states introduced in \cite{Valle_fermionic_coherent_states_in_a_fock_superspace} which will have the same form as the states in Eq.\eqref{grassman coherent states}, $\mathbf{c}^{\dagger},\mathbf{c}$ acting on $\mathcal{H}_F=\mathbb{C}^2$ in the same way as $\mathbf{J}_+, \mathbf{J}_-$ act on $\mathcal{H}_R=\mathbb{C}^2$. Other families of coherent states associated with supermanifolds could be obtained by means of representations of super Lie groups on some Hilbert space. For a detailed exposition of the theory of super Lie groups and their representations one could refer, for instance, to \cite{carmeli_cassinelli_toigo_varadarajan_unitary_representations_of_super_lie_groups}.

\section{Conclusions}
In this paper we have investigated the possibility of inducing nonlinear dynamics on submanifolds $\mathcal{M}$ of the Hilbert space $\mathcal{H}$ of a chosen quantum system. In the first part we mainly analysed invariant submanifolds with respect to specific unitary evolutions. The quantum evolution induces a one-parameter group of transformations on $\mathcal{M}$ and we have illustrated an example where this dynamics is nonlinear. Furthermore we have shown that, even if the embedding of $\mathcal{M}$ into $\mathcal{H}$ is nonlinear, the induced evolution can be linear, and this happens for the immersions defined via Radcliffe coherent states, and for the bosonic and fermionic oscillators.

On the other hand, the second part is centered on a different procedure to induce dynamical evolutions on immersed submanifolds of $\mathcal{H}$. The main ingredient is the introduction of a Lagrangian function $\lag$ on $T\mathcal{H}_0$ for the Schr\"{o}dinger equation. Then, the pullback of this Lagrangian via the immersion map always exists and defines a new Lagrangian function $\tilde{\lag}$ on $T\mathcal{M}$ and all the machinery of Lagrangian mechanics can be applied to obtain equations of motion on $\mathcal{M}$. Even if in this paper we have presented examples in which the immersed submanifold $\mathcal{M}$ is invariant under the unitary evolution associated with Schr\"{o}dinger equation, these are exceptions rather than a typical situation. A more detailed analysis of the possible induced dynamics is the aim of future works. In particular it would be interesting to investigate the possibility of defining approximation procedures via this Lagrangian immersion, in a way similar to the variational method for obtaining an approximation of the ground states of Hamiltonian operators. Analogously, one could think of the solution of the induced Lagrangian problem on a finite dimensional $\mathcal{M}$ as an approximation of the quantum evolution on an infintite dimensional Hilbert space. 

Finally, in this paper we have considered only quantum unitary evolution. The problem of classical-like dynamics induced by dynamics of some quantum open system will be addressed elsewhere. However, as a concluding remark, we want to give a simple example showing how this procedure could be performed. Let us consider the example of a quantum dynamics and its reduction for a qubit state obeying to the GKLS equation studied in \cite{Wudarski_Chruscinski_Markovian_semigroups_from_non_Markovian_evolutions}.
The non-Markovian evolution of the density matrix reads:
\be
\rho_{0}\mapsto \rho_{t}= \frac{1 + \mathrm{e}^{-ct}}{2} \rho_{0} + \frac{1 + \mathrm{e}^{-ct}}{4}\left(\sigma_{1}\rho_{0}\sigma_{1} + \sigma_{2}\rho_{0}\sigma_{2}\right)\,. 
\ee
For:
\be
\rho_{0}=\left(\begin{matrix}\rho_{11} & \rho_{12} \\ \rho_{12}^{*} & 1 - \rho_{11} \end{matrix}\right)\,,
\ee
one gets the following expression for the evolution:
\be
\rho_{11}(t)=\rho_{11}\,\mathrm{e}^{-ct}\,,\;\;\;\rho_{12}(t)= \rho_{12} \frac{1 + \mathrm{e}^{-ct}}{2}\,.
\ee
In terms of the Bloch parameters:
\be
\rho_{0}=\frac{1}{2}\left(\sigma_{0} + x\,\sigma_{1} + y\,\sigma_{2} + z\,\sigma_{3}\right)\,,
\ee
we have:
\be
\begin{split}
x(t)& =\frac{x}{2}\left(1 + \mathrm{e}^{-ct}\right) \\
y(t)& =\frac{y}{2}\left(1 + \mathrm{e}^{-ct}\right) \\
z(t)& =z\, \mathrm{e}^{-ct}\,.
\end{split}
\ee
It is clear that the equatorial disc is a subset of the space of quantum states of the qubit (Bloch ball) which is invariant with respect to the dynamical evolution considered.
If we consider the manifold (with corners):
\be
\mathcal{M}:=\left\{(r\,,\theta)\in\mathbb{R}^{2}\colon 0\leq r\leq R\leq 1\,, 0\leq\theta\leq\Theta<2\pi \right\}
\ee
and the injective immersion of this manifold into the space of quantum states of the qubit (Bloch ball) given by:
\be
(r\,,\theta)\mapsto \rho\left\{\begin{matrix} x=r\,\cos(\theta) \\ y= r\,\sin(\theta) \\ z=0 \end{matrix}\right.\,,
\ee
it is a matter of straightforward calculation to show that $i(\mathcal{M})$ is invariant with respect to the dynamical evolution given above.
This means that the quantum non-Markovian dynamics induces a classical-like dynamical evolution on the manifold (with corner) $\mathcal{M}$.
Consequently, if we immerse an abstract disc $D$ into the space of quantum states of the qubit by identifying elements of $D$ with the quantum states in the equatorial disc of the Bloch ball, the non-Markovian quantum dynamics induces a non-Markovian classical-like dynamics on the disk $D$.
We may also consider a three dimensional invariant manifold by considering any cylinder built on a disc lying in the equatorial plane of the Bloch ball.

\bigskip

\noindent{\bf Acknowledgements} 
P.V.  acknowledges  support by COST (European Cooperation in Science  and  Technology)  in  the  framework  of  COST  Action  MP1405  QSPACE. GM would like to thank the support provided by the Santander/UC3M Excellence Chair Programme 2019/2020;he also  acknowledges financial support from the Spanish Ministry of Economy and Competitiveness, through the Severo Ochoa Programme for Centres of Excellence in RD(SEV-2015/0554).G. M. is a member of the Gruppo Nazionale di Fisica Matematica (INDAM),Italy. F.D.C. would like to thank partial support provided by the MINECO research project MTM2017-84098-P and QUITEMAD++, S2018/TCS-A4342.

\end{document}